\documentclass[preprint2]{aastex}

\def\cs{{c_{\rm s}}}
\newcommand{\dfrac}[2]{\frac{\displaystyle{#1}}{\displaystyle{#2}}}

\shorttitle{Molecular Hydrogen Emission from PPDs}
\shortauthors{Nomura et al.}

\begin{document}


\title{Molecular Hydrogen Emission from Protoplanetary Disks II. \\
Effects of X-ray Irradiation and Dust Evolution}


\author{H. Nomura\altaffilmark{1,3}, Y. Aikawa\altaffilmark{1}, M. Tsujimoto\altaffilmark{2}, Y. Nakagawa\altaffilmark{1}, and T.J. Millar\altaffilmark{3}}

\email{h.nomura@qub.ac.uk}


\altaffiltext{1}{Department of Earth and Planetary Sciences, 
Kobe University, 1-1 Rokkodai-cho, Nada, Kobe 657-8501, Japan}
\altaffiltext{2}{Department of Physics, Rikkyo University, 3-34-1
Nishi-Ikebukuro, Toshima, Tokyo 171-8501, Japan} 
\altaffiltext{3}{Astrophysics Research Centre, School of Mathematics and Physics,
Queen's University Belfast, Belfast BT7 1NN, Northern Ireland, UK}


\begin{abstract}

Detailed models for the density and temperature profiles of gas and dust
 in protoplanetary disks are constructed by taking into account X-ray
 and ultraviolet (UV) irradiation from a central T Tauri star, as well
 as dust size growth and settling toward the disk midplane. 
The spatial and size distributions of dust grains in the disks are
 numerically computed by solving the coagulation equation for settling dust
 particles, with the result that the mass and total surface area of dust
 grains per unit volume of the gas in protoplanetary disks are very small,
 except at the disk midplane. 
 The level populations and line emission of molecular hydrogen are
 calculated using the derived physical structure of the disks. 
X-ray irradiation is the dominant
 heating source of the 
 gas in the inner disk region and in the surface layer, while
 the far UV heating dominates otherwise.
If the central star has strong X-ray and weak UV radiation, the level
 populations of molecular hydrogen are controlled by X-ray pumping, 
and the X-ray induced transition lines could be observable.
If the UV irradiation is strong, the level populations are controlled
 by thermal collisions or UV pumping, depending on the
 properties of the dust grains in the disks.
As the dust particles evolve in the disks, the gas temperature at the
 disk surface drops because the grain photoelectric heating becomes less
 efficient, while the UV radiation fields become stronger due to the
 decrease 
 of grain opacity. This makes the level populations of molecular hydrogen
 change from local thermodynamic equilibrium (LTE) to non-LTE
 distributions, which results in changes to the line ratios of molecular
 hydrogen emission. Our results suggest that dust evolution in 
 protoplanetary disks could be observable through the line ratios of
 molecular hydrogen.
 The emission lines are strong from disks irradiated by strong
 UV and X-rays and possessing small dust grains; such disks will be good
 targets in which to observe molecular hydrogen emission. 
\end{abstract}




\section{Introduction}

Observations of thermal dust continuum emission (e.g., Kenyon \&
Hartmann 1987; Beckwith \& Sargent 1993; Kitamura et al. 2002) and, more
directly, images of light scattered from dust grains 
(e.g., Roddier et al. 1996; Itoh et al. 2003; Duchene et al. 2004)
have revealed that young stellar objects have circumstellar disks.
In addition, a gaseous component has been
detected from the circumstellar disks around T Tauri stars (e.g., Carr
1989; Dutrey et al. 1997, 2007; Thi et al. 2004; Qi et al. 2006; Najita
et al. 2007; Bergin et al. 2007). Recent high spectral
resolution and high sensitivity observations have made it possible to
detect line emission of molecular hydrogen gas, which is the major
component of the gas in the protoplanetary disks (Thi et al. 1999,
2001a, b; Richter et al. 2002; Sheret et al. 2003; Sako et al. 2005;
Weintraub et al. 2000; Bary et al. 2002, 2003; Itoh et al. 2003;
Herczeg et al. 2002, 2004, 2006; Bergin et al. 2004).
Furthermore, detection of molecular hydrogen lines in the near- and
mid-infrared 
wavelength bands has been reported towards T
Tauri stars and Herbig Ae/Be stars very recently
(Weinstraub et al. 2005; Bary et al. 2007, in preparation; Richter et
al. 2007; Bitner et al. 2007, in preparation).
In a previous paper (Nomura \& Millar 2005, hereafter Paper I), we
constructed a model for molecular hydrogen emission from a
protoplanetary disk that is irradiated by strong ultraviolet (UV)
radiation from a central star, and whose dust component has the same
properties as dense molecular cloud dust.

Now, it is known observationally that many young stellar objects
emit strong X-ray radiation (Koyama et al. 1994; Feigelson \&
Montmerle 1999; Tsujimoto et al. 2002; Imanishi et al. 2003; Getman et
al. 2005; Preibisch et al. 2005) which ionizes hydrogen gas and
could be one 
of the important heating sources of gas in protoplanetary disks,
in addition to the grain photoelectric
heating induced by far UV radiation from the central star (cf. Kamp \&
Dullemond 2004; Jonkheid et al. 2004; Dullemond et al. 2007;
Paper I). Actually,
some model calculations show that X-ray irradiation can heat the gas
to very high temperatures at the surface layer of the disks 
(Glassgold \& Najita 2001; Gorti
\& Hollenbach 2004; Glassgold et al. 2004; Kamp et al. 2005). 
Furthermore, it has been suggested that fast secondary electrons
produced by X-ray ionization, similar to the electrons induced by
cosmic-ray ionization, can pump molecular hydrogen into excited
electronic states and may be important, for example, in extra-galactic 
objects, Herbig-Haro objects, and supernova remnants 
(e.g., Shemansky et al. 1985; Gredel et al. 1989; Gredel \& Dalgarno
1995; Tin\'e et al. 1997). This X-ray pumping could be observable
toward disks which are irradiated by strong X-ray
radiation from their central stars (e.g., Bergin et al. 2004).

As disks evolve, it is believed that the dust particles in the disks coagulate
and settle toward the disk midplane as the first step of planet
formation. The dust dynamics in this stage have been studied
theoretically in many works (e.g., Weidenschilling 1980, 1997; Nakagawa
et al. 1981, 1986; Mizuno et al. 1988; Mizuno 1989; Cuzzi et al. 1993;
Schmitt et al. 1997; Nomura \& Nakagawa 2006). This dust evolution is
expected to affect observational properties of the disks, and some
model calculations have been done in order to study the effect (e.g.,
Miyake \& Nakagawa 1993, 1995; D'Alessio et al. 2001, 2006; Dullemond \&
Dominik 2004; Jonkheid et al. 2004, 2006, 2007; Retting et al. 2006;
Aikawa \& Nomura 2006). In
addition, some numerical calculations of the dust evolution have been
done by solving the coagulation equation for settling dust particles,
and its effect on the spectral energy distribution of thermal dust
emission disks investigated (Suttner \&
Yorke 2001; Tanaka et al. 2005; Dullemond \& Dominik 2005).
In this paper, we further examine the effects of the dust evolution on
the physical structure of the gas in the disks and on molecular hydrogen
emission by using both a simple dust model and a
numerical calculation of the coagulation equation.

Historically, line emission from molecular hydrogen has been
observed towards various kinds of astronomical objects, such as shock
fronts associated with star forming regions, reflection nebulae,
planetary nebulae, supernova remnants, external galaxies, and so on
(e.g., Beckwith et al. 1978; Brown et al. 1983; Hasegawa et al. 1987;
Burton et al. 1992). 
The observed line ratios probe the physical properties of these
objects as they reflect the
excitation mechanisms of molecular hydrogen, e.g., thermal collisions,
ultraviolet and X-ray pumping, and formation pumping 
(e.g., Black \& van Dishoeck 1987;
Sternberg \& Dalgalno 1989; Tanaka et al. 1989; Tin\'e et al. 1997;
Takahashi \& Uehara 2001). 
In this work we propose a possible
observational diagnostic of the dust evolution in protoplanetary disks 
using line spectra and the line ratios of molecular hydrogen.

In the following sections, we model the density and temperature
profiles of the gas and dust in protoplanetary disks, taking into
account the X-ray and UV irradiation from a central star, as well as 
dust growth and settling towards the disk midplane. 
Then, using the physical structure, we calculate the level populations and
line emission of molecular hydrogen. 
In \S 2,
we introduce the models we use in this work. For the dust evolution,
we use both a simple model and a more realistic model in which we
solve the coagulation equation for settling dust particles (\S\ref{S2.1}).
The X-ray and UV radiation fields are computed 1+1 dimensionally
(\S\ref{S2.2}), the density and temperature profiles of the gas and dust
in the disks are obtained by assuming vertical hydrostatic equilibrium
and local thermal and radiative equilibrium (\S\ref{S2.3}). The level
populations of molecular hydrogen are calculated under an assumption of
statistical equilibrium from which we get the molecular
hydrogen emission by solving the radiative transfer equation (\S\ref{S2.4}).
In \S 3, we present the resulting dust size and spatial distributions
(\S\ref{S3.1}), the 
physical structure of the disks (\S\ref{S3.2}), the level populations of
molecular hydrogen (\S\ref{S3.3}), and the line spectra and line ratios
of molecular hydrogen (\S\ref{S3.4}), in which the effects of
the X-ray irradiation and the dust evolution are discussed.
Finally, the results are summarized in \S 4.

\section{Models}

\subsection{Spatial and Size Distributions of Dust Particles}\label{S2.1}

The physical structure and the molecular hydrogen emission of the disks
are affected by the dust model in various ways; for example, the UV radiation
field through dust extinction (\S\ref{S2.2}), the dust temperature through
optical properties of dust grains, the gas temperature through grain
photoelectric heating 
and dust-gas collisions (\S\ref{S2.3}), and the molecular hydrogen formation
rate on dust grains (\S\ref{S2.4}). 
In this paper we use the following two types of model for the dust size
and spatial distributions. In model A we adopt a very simple assumption
in order to understand the basic properties of the effects of dust evolution.
In model B we consider a more realistic case by numerically solving the 
coagulation equation for settling dust particles.

In both models the shape of the dust particles is simply assumed to be a
compact sphere (see e.g., Kozasa et al. 1992; Ossenkopf
1993; Ormel et al. 2007 for fractal dust aggregate models). Making use
of the resulting 
spatial and size distributions of dust grains, the dust absorption
($\kappa_{\nu}$) and scattering ($\sigma_{\nu}$) coefficients at each
position of the disk are computed by means of 
the Mie theory (Bohren \& Huffman 1983), with the dust particles
assumed to consist of silicate, carboneous grains, and water ice (see
Paper I for details). 

\subsubsection{Model A} 

In model A we assume that
the dust grains have a spatially uniform distribution and the mass
fractional abundance of the dust with respect to the gas is
fixed at each position in the disk (i.e. the dust grains are well mixed with
the gas). The size distribution is set to be 
$dn/da\propto a^{-3.5}$ ($a$ is the radius of dust grain) with the
maximum grain radii of $a_{\rm max}=10\mu$m, 1mm, and 10cm. The minimum 
radius is set to be $0.01\micron$ for all models. 
The amount of small dust grains decreases with increasing maximum
grain radii as we keep the mass fractional abundance of the dust
grains to the gas fixed (see
also \S 2.1.3) (e.g., Miyake \&
Nakagawa 1993; D'Alessio et al. 2001; Aikawa \& Nomura 2006).  

\subsubsection{Model B}\label{S2.1.2}

In this model the spatial and size distributions of dust grains are
obtained by solving coagulation equations for various sizes of settling
dust particles,
\begin{displaymath}
\dfrac{\partial\varphi(i)}{\partial t}+\dfrac{\partial}{\partial z}[V_z(i)\varphi(i)]=-m_i\varphi(i)\sum_{j=1}^n\beta(i,j)\varphi(j)
\end{displaymath}
\begin{equation}
+\dfrac{1}{2}m_i\sum_{j=1}^{i-1}\beta(i-j,j)\varphi(i-j)\varphi(j),  \label{eq.2-1}
\end{equation}
where $\varphi(i)$ is the mass density of dust particles in a mass bin
$i$ whose
summation is equal to the total mass density of the dust particles at a
given position and time as
\begin{equation}
\rho_{\rm dust}(x,z,t)=\sum_{i=1}^n\varphi(x,z,t,i). \label{eq.2-2}
\end{equation}
Here we briefly summarize the dust evolution model; more details can be 
found in Nomura \& Nakagawa (2006). Now, $V_z(i)$ in equation 
(\ref{eq.2-1}) is the vertical velocity
and $m_i$ is the typical mass of a particle in a mass bin $i$. 
%
The symbol $\beta(i,j)$ is related to the sticking rate of two colliding
dust particles, given by
\begin{equation}
\beta(i,j)=\pi(a_i+a_j)^2\delta Vp_{\rm s}/m_im_j, 
\end{equation}
where $a_i$ is the radius of a dust particle in a mass bin $i$, and we
simply assume the sticking probability of $p_{\rm s}=1$ in this paper. 
The fragmentation of dust particles is simply neglected in this
work.  
For the relative velocity between two colliding particles, $\delta V$,
we adopt  
\begin{equation}
\delta V=(\delta V_{\rm B}^2+\delta V_z^2+\delta V_x^2+\delta V_{\rm T}^2)^{1/2}.
\end{equation}
The symbol $\delta V_{\rm B}=(8kT_d/\pi)^{1/2}(1/m_i+1/m_j)^{1/2}$
($k$ is Boltzmann's constant and $T_d$ is the dust temperature) is the
relative velocity caused by the thermal Brownian motion. The symbols
$\delta 
V_z=V_z(i)-V_z(j)$ and $\delta V_x=V_x(i)-V_x(j)$ ($V_x(i)$ is the local
radial velocity component of dust particles with mass $m_i$, arising
from angular momentum loss via gas-dust friction) are the velocity
differences in the vertical and radial directions, respectively. 
Finally, $\delta V_{\rm T}$ is the turbulence-induced  
relative velocity (see Nomura \& Nakagawa 2006 for more details). 
The disk is simply assumed to be completely quiescent or
turbulent, with $\delta V_{\rm T}=0$ in a quiescent disk model.

The mass flux in equation (\ref{eq.2-1}) is given by
\begin{equation}
V_z(i)\varphi(i)=-\dfrac{\Omega_{\rm K}^2z}{A\rho}\varphi(i)
\end{equation}
in a quiescent disk, where $\Omega_{\rm K}$ is the Keplerian
frequency and $\rho$ is the gas density.
For the drag coefficient between the gas and dust particles, $A$, we
adopt $A=\cs/\rho_sa$ for $a\la l_g$ and $A=3\cs l_g/2\rho_sa^2$ for
$a\ga l_g$, following Epstein's and Stokes' laws, respectively, where
$\cs$, $\rho_s$, and $l_g$ are the sound speed of the gas, the solid
density of a dust particle, and the mean free path of gas particles,
respectively. The mean velocity of the dust particles
in the vertical direction, $V_z(i)=\Omega_{\rm K}^2z/A\rho$, is obtained by 
balancing the gas-dust friction force, $A\rho V_z(i)$, and the
gravitational force in the vertical direction, $\Omega_{\rm K}^2z$. In a
turbulent disk, the mass flux is written as 
\begin{equation}
V_z(i)\varphi(i)=-\dfrac{\Omega_{\rm K}^2z}{A\rho}\varphi(i)-D_0\rho\dfrac{\partial[\varphi(i)/\rho]}{\partial z},
\end{equation}
following the gradient diffusion hypothesis.
For the turbulent diffusivity, we adopt $D_0=\alpha'\cs
H/(1+\Omega_{\rm K}/A\rho)$, where $H=\cs_0/\Omega_{\rm K}$ ($\cs_0$ is
the sound speed at the disk midplane) is the scale height of the
disk and we set $\alpha'=10^{-4}$ in this paper. 

The global radial motion of the dust particles toward the central star
is not taken into account in this work. 
At the disk surface this simplified treatment is applicable because
large dust particles settle toward the disk midplane more rapidly than
they move toward the central star (Figs.~\ref{f3} and \ref{f4} in
\S\ref{S3.1} actually show that large particles cannot stay in the
surface layer). 
We note that the radial motion is negligible for the small dust
particles which couple with the gas efficiently via friction force 
(e.g., Adachi et al. 1976; Weidenschilling 1977; Takeuchi \& Lin 2005).
Therefore, neglect of radial global motion will not affect our results
of molecular hydrogen lines which are mainly emitted in the disk surface
(see Paper I). In a completely quiescent disk the radial motion is
negligible all over the disk (e.g., Nakagawa et al. 1986).
Now, in order to avoid making very large particles which do not couple
with the gas and should fall on towards the central star, we remove
particles larger than some critical radius 
by simply assuming that  
when the dust particles grow large enough so that they cannot be trapped in
a turbulent eddy, they gain very rapid radial motion. The critical radius
is estimated as $a_{\rm crit}=\cs\rho_{\rm gas}/\rho_s\Omega_{\rm K}$
for $a\la l_g$ and $a_{\rm crit}=(3\cs\rho_{\rm
gas}l_g/2\rho_s\Omega_{\rm K})^{1/2}$ for $a\ga l_g$ by comparing the
friction time between the gas and dust particles, $\tau_f=1/A\rho$, and
the turnover time of the largest turbulent eddy, $\tau_{\rm
eddy}=1/\Omega_{\rm K}$. We note that the dust particles grow to be
larger than $a_{\rm crit}$ only
close to the midplane of a turbulent disk (see \S\ref{S3.1}).
%

As the initial condition of the calculations, we set the dust particles
to be well-mixed with the gas and the dust-to-gas mass ratio to be
spatially uniform.
For the initial size distribution of the dust grains, we adopt
dust model A with $a_{\rm max}=10\micron$, which is similar to the
dust model of 
dense molecular clouds (e.g., Weingartner \& Draine 2001). We also
assume that the disk is surrounded by a dense molecular cloud with gas
number density $n_{\rm out}=10^4$cm$^{-3}$, and consider
a continuous input of dust particles from the molecular clouds to the
disk 
as a boundary condition at the disk surface ($z=z_{\rm coag}$, see
below and \S\ref{S3.1}) for the calculations of dust evolution
(Eq. [\ref{eq.2-1}]). 
Dust particles fall on to the disk because the pressure
gradient force is negligibly small for them (e.g., Landau et al. 1967)
and cannot sustain them against the gravitational force of the central
star (we note that the gas is assumed to be stationary due to the
pressure gradient force).
This input of dust particles has great influence on structure of
the disk surface and observational properties of the disk (see
\S\ref{S3}).
The total mass of dust particles infalling from the cloud to the disk
in $10^6$ yr (the time for which the calculations are performed) is 
$\sim 5\times 10^{-5}M_{\odot}$, which corresponds to a dust mass in a
spherical cloud with a radius of 2,500 AU and a gas density of $10^4$cm$^{-3}$,
and $1/3$ of the initial mass of dust grains in the disk.
In the calculations for the dust coagulation and settling
we set 23 radial grids
logarithmically for $x=0.2-100$ AU, and 50 vertical grids for the region
where the coagulation becomes significant ($z<z_{\rm coag}$, see
\S\ref{S3.1}). 
%
%
The dust coagulation equation is solved using fixed density and
temperature profiles of dust and gas because self-consistent
calculations coupled to the time evolution of the dust and gas profiles
are very time-consuming. Instead, 
we get the dust distribution and the temperature and density profiles 
by iterating the calculations only once under an assumption
that the temperature and density profiles turn into an equilibrium state
very quickly; that is, we (a) calculate the dust
gas profiles using the initial dust distribution, (b) solve the
coagulation equation using the temperature and density profile in (a),
(c) compute the dust 
and gas profiles again using the dust distribution in (b), (d) obtain
the dust distribution by solving the coagulation equation with the
temperature and density profiles in (c), and (e) finally
get the dust and gas profiles using the dust distribution in (d).
%
In the calculations in \S\ref{S3} we compare the gas temperature and
density profiles in (c) and (e), and checked that the errors are
within 30\% for the gas temperatures and 80\% at most for the gas
densities in both quiescent and turbulent disks. These errors are
relatively large at the upper layer of the disks where 
the molecular hydrogen is photodissociated via UV radiation from the
central star, so the errors in molecular hydrogen emission from the disks
(see \S\ref{S3.4}) are smaller and within 15\% for all lines.

\subsubsection{Parameter for Total Surface Area of Dust Particles, $f_{\rm dust}$}\label{S2.1.3}

We introduce a parameter, $f_{\rm dust}$, that represents the total
surface area of dust particles per unit volume of the gas at each
position in the disk $(x,z)$, 
\begin{equation}
f_{\rm dust}(x,z)=A_{\rm tot}(x,z)/A_{{\rm tot}, 0}, \label{eq7}
\end{equation}
where
\begin{equation}
A_{\rm tot}(x,z)=\int 4\pi a^2\dfrac{dn(x,z)}{da}da.
\end{equation}
The size distributions $dn/da\propto\varphi(i)/a_i^4$ in dust model
B, and $A_{{\rm tot}, 
0}$ is calculated using the dust model A with $a_{\rm max}=10\mu$m,
which is similar to the dust model of dense molecular clouds and is used
as the initial condition for dust model B. This parameter controls the
physical disk structure and molecular hydrogen emission from the disks
because the grain opacity, the grain photoelectric heating rate, and the
formation rate of molecular hydrogen on grain surface are roughly
proportional to it. Thus, the dust and gas temperatures, the
UV radiation field, and the abundance of molecular hydrogen are related
to $f_{\rm dust}$ (see the following subsections). In dust model A,
the parameter $f_{\rm dust}$ has values of 1, 0.1, and 0.01 for the maximum
dust size of $a_{\rm max}=10\mu$m, 1mm, and 10cm, respectively. 
In model B, $f_{\rm dust}$ decreases with dust
size growth and settling (except at the disk midplane), which leads to
a decrease of the grain photoelectric heating rate, and thus the gas
temperature and so on (see \S\ref{S3.1} for more details).


\subsection{X-ray and Ultraviolet Radiation Fields}\label{S2.2}

\begin{figure}[t]
\includegraphics[scale=1.0]{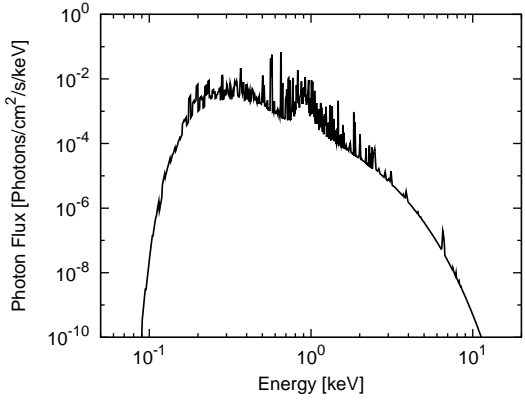}
\caption{The model spectra of the X-ray radiation at the central star,
 which reproduces observation toward a classical T Tauri star, TW
 Hya ($d=56$pc). \label{f1}}
\end{figure}

Observations have shown that many T Tauri stars emit strong X-ray
(e.g., Koyama et al. 1994; Feigelson \& Montmerle 1999; Tsujimoto et
al. 2002; Imanishi et al. 2003; Getman et al. 2005; Preibisch et
al. 2005) as well as strong ultraviolet (UV) radiation (e.g., Herbig \&
Goodrich 1986; Herbst et al. 1994; Valenti et al. 2000). 
For the X-ray radiation from
the central star we use a model which reproduces observational data
toward a classical T Tauri star, TW Hydrae
(cf. Kastner et al. 2002; Stelzer \& Schmitt 2004).
Retrieving the archived \textit{XMM-Newton} data, we fit the
spectrum with a two-temperature thin-thermal plasma model (mekal model; 
Mewe et al. 1985; Kaastra et al. 1992; Liedahl et al. 1995) which is
often used in order to reproduce observed X-ray spectrum of T Tauri
stars. The derived best-fit parameters are $kT_1=0.8$keV and
$kT_2=0.2$keV for the plasma temperatures, and $N_{\rm H}=2.7\times
10^{20}$ cm$^{-2}$ for the foreground interstellar hydrogen column
density.
The total X-ray luminosity of the spectrum
corresponds to $L_X\sim 10^{30}$ erg s$^{-1}$.
In Figure~\ref{f1} the resulting model spectra is plotted.
The adopted stellar UV radiation field model is also based on
observations towards TW Hydrae and analyses by Herbst et al. (1994),
Costa et al. (2000), Bergin et al. (2003), Herczeg et al. (2002), and
Ardila et al. (2002). The model consists of photospheric black 
body radiation, hydrogenic thermal bremsstrahlung radiation, and strong 
Ly $\alpha$ line emission (see Appendix C of Paper I). The total FUV
(6eV $<h\nu <$ 13eV) luminosity corresponds to $L_{\rm FUV}\sim 10^{31}$
erg s$^{-1}$. The interstellar UV radiation field is
taken into account, but its contribution is negligible under the
strong UV irradiation from the central star (see Paper I for details
of the UV radiation field in the disk). 
We note that although we use the X-ray and UV radiation of TW Hya
(as it is one of the most well-observed T Tauri stars), our disk model
is to be more widely applicable.

The X-ray and UV fields in the disk are calculated in 1+1 dimensions
in the radial and vertical directions (see also Paper I) as
\begin{equation}
F_{\nu, R}(R,\Theta)=fF_{\nu, {\rm star}}\exp(-\tau_{\nu, R}),\ \ \ \tau_{\nu, R}=\int_{R_*}^R \chi_{\nu}\rho dR, \label{eq.2-6}
\end{equation}
and
\begin{displaymath}
F_{\nu, z}(x,z)=F_{\nu, {\rm ISRF}}\exp[-\tau_{\nu, z}(z_{\infty})]
\end{displaymath}
\begin{displaymath}
+2\pi\int_z^{z_{\infty}} \sigma_{\nu}(x,z')\rho(x,z')F_{\nu, R}(x,z')e^{-\tau_{\nu, z}(z')}dz',
\end{displaymath}
\begin{equation}
\tau_{\nu, z}(z')=\int_z^{z'} \chi_{\nu}\rho dz'', \label{eq.2-7}
\end{equation}
where the direct radiation from the central star is calculated
fully, while a plane-parallel approximation in the vertical direction is 
adopted for the calculation of the scattering process, which
could result in overestimating the radiation fields in the disks.
In the above equations 
$F_{\nu, {\rm star}}$ is the specific radiation field at the
stellar surface and  $f=(R_*/R)^2$ accounts for the geometrical 
dilution of the radiation field. 
$F_{\nu, {\rm ISRF}}$ is the
FUV interstellar radiation field, and we set $F_{{\rm Xray,\ ISRF}}=0$
in the calculation of the X-ray radiation field. $\tau_{\nu, R}$ and
$\tau_{\nu, z}$ are the specific optical depths from the stellar
surface $(R_*,\Theta)$ to a point $(R,\Theta)$ and from a point $(x,z)$
to $(x,z')$, respectively. $\rho$ is the gas density, and $\chi_{\nu}$
is the monochromatic extinction coefficient defined by the absorption
($\kappa_{\nu}$) and scattering ($\sigma_{\nu}$) coefficients as
$\chi_{\nu}\equiv\kappa_{\nu}+\sigma_{\nu}$. 
%
In order to treat X-ray extinction, we adopt the attenuation
cross section at an energy $E$
of $\sigma_{\rm att}(E)=\sigma_{\rm ph}(E)+\sigma_{\rm Com}(E)$, where
$\sigma_{\rm ph}$ is the total photoionization cross section due to all
elements per hydrogen nucleus 
and $\sigma_{\rm
Com}$ is the incoherent Compton scattering cross section of hydrogen. 
For the cross section $\sigma_{\rm ph}$ we adopt a broken power-law
model given by Maloney et al. (1996; see also Wilms et al. 2000), and
$\sigma_{\rm Com}$ is calculated based on McMaster et
al. (1969, http://cars9.uchicago.edu/\textasciitilde newville/mcbook/). 
In calculating equations (\ref{eq.2-6}) and (\ref{eq.2-7}), we adopt
$\chi_{\rm Xray}=\sigma_{\rm att}(E)/m_{\rm p}$ and $\sigma_{\rm Xray}=\sigma_{\rm
Com}(E)/m_{\rm p}$, where $m_{\rm p}$ is the proton mass.


\subsection{Physical Structure of the Disks}\label{S2.3}

We model an axisymmetric disk surrounding a central star with the
physical parameters of typical T Tauri stars; a mass of
$M_*=0.5M_{\odot}$, a radius of $R_*=2R_{\odot}$, and a temperature
of $T_*=4000$K (e.g., Kenyon \& Hartmann 1995).

The gas temperature and density distributions of the disk are obtained
self-consistently by iteratively solving the equations for hydrostatic
equilibrium in the vertical direction and local thermal balance between
heating and cooling of gas (see Paper I for details).
The vertical hydrostatic equilibrium is represented by an equation, 
\begin{equation}
\dfrac{dP}{dz}=-\rho\dfrac{GM_*z}{(x^2+z^2)^{3/2}}. \label{eq.1}
\end{equation}
$G$ is the gravitational constant, and $P$ is the gas pressure given by
$P=\rho kT/m_{\mu}$, where $\rho$, $T$, 
$k$, and $m_{\mu}$ are the density and temperature of the gas,
Boltzmann's constant, and the mean molecular mass, respectively. 
The condition, $\int_{-z_{\infty}}^{z_{\infty}}\rho(x,z)dz=\Sigma(x)$,
is imposed, where we set $\rho(x,z_{\infty})=5.0\times
10^{-19}$ g cm$^{-3}$ ($n_{\rm tot}\approx 3\times 10^5$cm$^{-3}$) as the
boundary condition. The surface density at a disk radius $x$,
$\Sigma(x)$, is defined by assuming a constantly accreting viscous disk
model and equating the gravitational energy release of accreting mass
to the thermal heating via viscous dissipation at the disk midplane,
\begin{equation}
\dfrac{9}{4}\Sigma\alpha\cs_0^2\Omega_{\rm K}=\dfrac{3GM_*\dot{M}}{4\pi x^3}
\biggl[1-\biggl(\dfrac{R_*}{x}\biggr)^{1/2}\biggr],
\label{eq.2}
\end{equation}
where $\cs_0$ and $\Omega_{\rm K}=(GM_*/x^3)^{1/2}$ represent the sound
speed at 
the midplane and the Keplerian frequency, respectively. A viscous
parameter of $\alpha=0.01$ and a constant mass accretion rate of
$\dot{M}=10^{-8}$ M$_{\odot}$ yr$^{-1}$ are adopted here.

The gas temperature, $T$, is obtained by assuming detailed
energy balance at each position in the disk,
\begin{equation}
\Gamma_{\rm FUV}+\Gamma_{\rm Xray}+L_{\rm gr}+\Lambda_{\rm line}=0, \label{eq.3}
\end{equation}
where we include grain photoelectric heating induced by
FUV photons, $\Gamma_{\rm FUV}$, X-ray heating caused by hydrogen
ionization, $\Gamma_{\rm Xray}$,
gas-grain collisions, $\Lambda_{\rm gr}$, and radiative
cooling by line transitions, $\Lambda_{\rm line}$ for the gas heating
and cooling processes. 
The X-ray heating rate, $\Gamma_{\rm Xray}$, is calculated as 
\begin{equation}
\Gamma_{\rm X ray}=n_{\rm tot}f_hH_{\rm X}, \label{eq.4}
\end{equation}
where $n_{\rm tot}$ is the total number density of hydrogen nuclei,
and $f_h$ is the heating efficiency, namely, the fraction of absorbed
energy that goes into heating the gas. We adopt $f_h=0.1$ for
atomic hydrogen and $f_h=0.4$ for molecular hydrogen (Maloney et
al. 1996; Gorti \& Hollenbach 2004). $H_{\rm X}$ is the
local X-ray energy deposition rate per particle, given by
\begin{equation}
H_{\rm X}=\int_{E_{\rm min}}^{E_{\rm max}} \sigma_{\rm ph}(E)F_{\rm X}(E)dE,
\label{eq.5}
\end{equation}
where $\sigma_{\rm ph}(E)$ is the total photoionization cross section
due to all elements per hydrogen nucleus at energy $E$.
The symbol $F_{\rm X}(E)$ is the X-ray energy flux at each
position in the disk, and $E_{\rm min}=0.1$keV and
$E_{\rm max}=10$keV are adopted for the minimum and maximum energy 
(see Fig.~\ref{f1} in \S\ref{S2.2}). 
We note that the viscous heating is not taken into account in the
energy balance (Eq. [\ref{eq.3}]) because it is not dominant (at the
disk surface) if $\alpha=0.01$ (Glassgold et al. 2004).

For radiative cooling by line transitions, we consider the Ly
$\alpha$ transition of atomic hydrogen and the metastable transition of
OI ($\lambda$ 6300\AA) in addition to the fine-structure transitions of
OI (63$\mu$m) and CII (158$\mu$m), and the rotational transitions of CO. 
In order to calculate the Ly $\alpha$ line cooling, we make use of the
table of level populations of atomic hydrogen for various electron
densities and temperatures, given by Storey \& Hummer (1995). The
collisional de-excitation rate coefficient is taken from Hollenbach \& 
McKee (1989) for calculation of OI $\lambda$ 6300 line cooling.
Paper I gives details of calculations of the OI and CII fine-structure,
and CO rotational transition line cooling.


The spatial and size distributions of dust grains affect the gas
temperature through
the grain photoelectric heating, $\Gamma_{\rm FUV}$, and the
energy exchange between gas and dust particles through collisions,
$\Lambda_{\rm gr}$. Both rates are
roughly proportional to the parameter which represents
the total surface area of the dust particles,
$f_{\rm dust}$, given in \S\ref{S2.1.3}. In this paper we simply set
$\Gamma_{\rm FUV}=f_{\rm dust}\Gamma_{{\rm FUV},0}$ and $\Lambda_{\rm
gr}=f_{\rm dust}\Lambda_{{\rm gr},0}$, 
where the heating/cooling rates with subscript '0' are calculated by
using the models given in Paper I in which we used the dense cloud dust
model (see also Aikawa \& Nomura 2006). 

The dust temperature
profile is important for determining the disk structure because the gas
temperature is well coupled to the dust temperature in the dense
region near the midplane of the disks. We obtain the dust temperature
by assuming local radiative equilibrium between absorption and
reemission of radiation by dust grains at each position in the disk.
The intensity is calculated by 
solving the axisymmetric two-dimensional radiative transfer equation
by means of the short characteristic method in spherical coordinates
(Dullemond \& Turolla 2000; Nomura 2002). As heating sources, we
consider the radiative flux produced by 
the viscous dissipation ($\alpha$-viscous model) at the disk midplane,
and the irradiation from the central star (see Paper I for details).
The dust evolution in the disks affects the dust temperature through the
change in grain opacity (\S\ref{S2.1}).

\subsection{Level Populations and Line Emission of Molecular Hydrogen}\label{S2.4}

In order to obtain the molecular hydrogen emission from the disk, we
first calculate the 
abundance and the level populations of the $X^1\Sigma_g^-$ electronic
state of molecular hydrogen in a statistical equilibrium state,
based on Wagenblast \& Hartquist (1988), as
\begin{displaymath}
n_l({\rm H}_2)\left[\sum_{m\ne l} \biggl(A_{lm}+\beta_{lm}+\gamma_{lm}+\sum_s n_sC_{lm}^s\biggr)+R_{{\rm diss},l}\right]
\end{displaymath}
\begin{displaymath}
+k_{{\rm O}+{\rm H}_2}n({\rm O})n_l({\rm H}_2)
\end{displaymath}
\begin{equation}
=\sum_{m\ne l}n_m({\rm H}_2)\biggl(A_{ml}+\beta_{ml}+\gamma_{ml}+\sum_s n_sC_{ml}^s\biggr)+n({\rm H})R_{{\rm form},l}, \label{eq.2.3.1}
\end{equation}
where $A_{lm}$ is the Einstein $A$-coefficient for spontaneous emission
from level $l$ to level $m$ and $C_{lm}^s$ is the collisional transition
rate with collision partner $s$. $\beta_{lm}$ represents the
effective rate for transition $l\rightarrow m$ via ultraviolet pumping
followed by radiative cascade, and $R_{{\rm diss},l}$ is the
photodissociation rate of hydrogen molecules in level $l$. $R_{{\rm
form},l}$ is the effective formation rate of H$_2$ in level $l$ on grain
surfaces. The endothermic reaction O + H$_2 \rightarrow$ OH + H, which 
destroys molecular hydrogen in high temperature regions, is also taken
into account (see Paper I for details). 
In addition, we consider the effective transition rate, $\gamma_{lm}$,
via X-ray pumping of molecular hydrogen followed by radiative cascade.
X-ray irradiation from the central star ionizes
the gas to produce photoelectrons and subsequently secondary electrons in
the disk. Through collisions they excite molecular hydrogen to singlet 
and triplet electronic states, followed by radiative
cascade down into the ground electronic state (e.g., Gredel \& Dalgarno
1995; Tin\'{e} et al. 1997; Bergin et al. 2004). In this paper we simply
use the entry efficiency $\alpha_{J_i}(v,J)$ from the levels $(v=0,J_i)$
to $(v,J)$ for fractional ionization of $10^{-4}$, tabulated in
Tin\'{e} et al. (1997), in order to estimate the 
rate $\gamma_{lm}$ as 
\begin{equation}
\gamma_{lm}=\zeta_{\rm X}\alpha_{J_l}(v_m,J_m). \label{eq.18}
\end{equation}
This simplified treatment will not cause significant error at the
middle layer of the outer disk where molecular hydrogen lines are mainly
emitted (the line fluxes are strong where the UV radiation field is not
too strong,
the gas temperature is moderately high, and the surface area is large;
see Paper I), but a full calculation of X-ray pumping and the
subsequent radiative cascade should be done in future. 
The symbol $\zeta_{\rm X}$ in equation (\ref{eq.18})
is the total hydrogen ionization rate given by
\begin{equation}
\zeta_{\rm X}\simeq N_{\rm sec}\int_{E_{\rm min}}^{E_{\rm max}} \sigma_{\rm ph}(E)F_{\rm X}(E)dE, \label{eq.2.4.3}
\end{equation}
where $\sigma_{\rm ph}$ is 
the total photoionization cross section
due to all elements per hydrogen nucleus,
and $N_{\rm sec}$ is the number of secondary
ionizations of hydrogen per unit energy produced by primary
photoelectrons and we put $N_{\rm sec}=26/$keV in this paper (e.g.,
Verner \& Yakovlev 1995; Maloney et al. 1996; Gorti \& Hollenbach 2004).
We note that the effect of interstellar cosmic-ray ionization
is not taken into account in the calculation of the pumping process,
but it will not affect the results as the X-ray
ionization rate is much higher at the disk surface (see \S\ref{S3.2.1}).
Possible reactions induced by X-rays are ignored, and a simple
chemical network given in Wagenblast \& Hartquist (1988) (plus the
reactions O + H$_2 \rightarrow$ OH + H and OH + $h\nu \rightarrow$ O +
H) is adopted as in Paper I. This 
neglect will not affect the resulting molecular hydrogen abundance since
the photodissociation by UV radiation from the central star or the
above-mentioned reaction with atomic oxygen is more efficient for
destroying molecular hydrogen than the X-ray-induced photoionization and
other related reactions in our model
(see e.g., Maloney et al. 1996 for XDR chemistry, and
also e.g., Aikawa \& Herbst 1999, 2001; Markwick et al. 2002 for more
detailed disk chemistry including the X-ray photoprocess).

The spatial and size distributions of dust grains affect the formation
rate of molecular hydrogen. The rate is roughly
proportional to the total surface area of the dust particles, $f_{\rm
dust}$, given in \S\ref{S2.1.3}, so we simply set $R_{{\rm form},l}=f_{\rm
dust}R_{{\rm form},l,0}$. Here we use the model in Paper I
in order to calculate $R_{{\rm form},l,0}$. 
The total formation rate is given by $\sum_l R_{{\rm
form},l}=7.5\times 10^{-18}f_{\rm dust}T^{0.5}\epsilon_{{\rm
H}_2}(T_d)n_{\rm tot}n(H)$ cm$^{-3}$ s$^{-1}$, where $T$ is the gas
temperature and $\epsilon_{{\rm H}_2}(T_d)$ 
is the recombination efficiency of atomic hydrogen on dust grains
as a function of 
the dust temperature, $T_d$ (Cazaux \& Tielens 2002,
2004; see also Pirronello et al. 1999; Zecho et al. 2002).

Making use of the physical properties obtained in the previous
subsections and the level populations, we calculate emission
(from levels $u$ to $l$) of molecular hydrogen from the disks
by integrating the radiative transfer equation (see Paper I for details), 
\begin{equation}
F_{ul}=\dfrac{1}{4\pi d^2}\int_{x_{\rm in}}^{x_{\rm out}}2\pi xdx\int_{-z_{\infty}}^{z_{\infty}}\tilde{\eta}_{ul}(x,z)dz, 
\end{equation}
where $\tilde{\eta}_{ul}(x,z)$ is the emissivity of the transition
line at $(x,z)$ times the effect of absorption in the upper disk layer,
given by
\begin{equation}
\tilde{\eta}_{ul}(x,z)=n_u(x,z)A_{ul}\dfrac{h\nu_{ul}}{4\pi}\exp(-\tau_{ul}(x,z)).
\end{equation}
$\tau_{ul}(x,z)$ is the optical depth from $z$ to the disk surface
$z_{\infty}$ at the frequency $\nu_{ul}$,
\begin{equation}
\tau_{ul}(x,z)=\int_z^{z_{\infty}}\chi_{ul}(x,z')dz', 
\end{equation}
where $\chi_{ul}$ is the total extinction coefficient,
\begin{equation}
\chi_{ul}=\rho\chi_{\nu_{ul}}+(n_lB_{lu}-n_uB_{ul})\Phi_{ul}\dfrac{h\nu_{ul}}{4\pi}.
\end{equation}
In these equations, $A_{ul}$ and $B_{ul}$ are the Einstein coefficients,
$n_u$ and $n_l$ are the number densities of the upper and lower levels,
respectively, and $\Phi_{ul}$ is the line profile function. The energy
difference between the levels $u$ and $l$ corresponds to $h\nu_{ul}$. 
The symbol $\chi_{\nu_{ul}}$ is the extinction coefficient of 
dust grains (see \S\ref{S2.1} and \S\ref{S2.2}) at the frequency
$\nu_{ul}$, and $\rho$ is the gas density.
Here, the disk is
assumed to be face on an observer, and we use the distance to an object
of $d=56$ pc for calculating the intensity in order to compare it with
the observations towards TW Hya.
Extinction by a foreground interstellar dust grains is not taken
into account in the calculations.

\section{Results}\label{S3}

\subsection{Spatial and Size Distributions of Dust Particles}\label{S3.1}

\begin{figure}[h]
\includegraphics[scale=1.0]{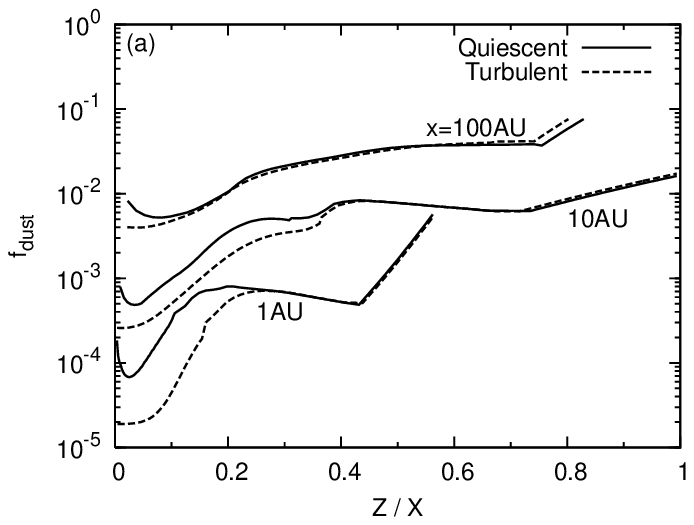}
\includegraphics[scale=1.0]{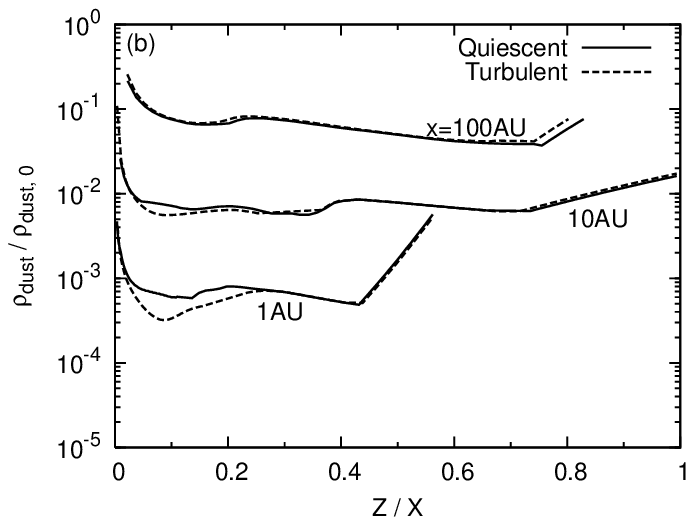}
\caption{The vertical profiles of (a) the parameter for the total
 surface area of dust grains, $f_{\rm dust}$, and (b) the total dust
 density, $\rho_{\rm dust}$, normalized by the initial value in
 quiescent ({\it solid lines}) and turbulent ({\it dashed lines}) disks
 at the disk radii of $x=1, 10$ and 100AU at $10^6$yr after the
 calculations start. At the disk surface 
 $\rho_{\rm dust}/\rho_{\rm dust, 0}$ and $f_{\rm dust}$ are small 
 due to the dust settling towards the disk midplane. Near the disk
 midplane $f_{\rm dust}$ is further smaller as a result of the
 dust coagulation, while $\rho_{\rm dust}/\rho_{\rm dust, 0}$ increases
 due to the dust settling. \label{f2}}
\end{figure}

For dust model B we obtain the spatial and size distributions of dust
particles by solving the coagulation equations for various sizes of
settling dust particles in a quiescent or turbulent disk (\S\ref{S2.1}).
In Figure \ref{f2} we plot the resulting profiles 
of (a) the parameter representing the total surface area of dust grains, 
$f_{\rm dust}$ (defined in Eq. [\ref{eq7}] in \S\ref{S2.1.3}), and (b) the
total dust density $\rho_{\rm dust}$ normalized by the initial value 
$\rho_{\rm dust, 0}$ (defined in Eq. [\ref{eq.2-2}] in \S\ref{S2.1.2}) 
in the vertical direction at the disk radii of $x=1, 10$, and 100AU. 
The initial dust density is simply proportional to the gas density (the
dust particles are well-mixed with the gas) and corresponds to
roughly 1\% of the gas mass density,
$\rho_{\rm dust, 0}\approx 0.01\rho$, in this model.
The solid and dashed lines are the profiles in quiescent and turbulent
disks, respectively. The calculations are performed for $10^6$
yrs, comparable to the typical age of classical T Tauri stars.
We note that at $t\sim 10^6$yr the dust coagulation process and settling
motion (input from the upstream and output to the downstream) are almost
in equilibrium state at each position in the disk, and the spatial and
size distributions of dust particles do not change with time
except in the region very close to the disk midplane.
Therefore, the spatial and size distributions of dust particles
in the surface layer presented in this subsection are applicable to
older star-disk systems as well.
Figure \ref{f2} shows that the mass and total surface area of dust
grains per unit volume of the gas
are much smaller than the initial values.
In the disk surface ($z>z_{\rm coag}$; see below)
where the density of particles is low enough so that
the dust particles settle before they grow, $\rho_{\rm dust}/\rho_{\rm
dust, 0}$ and $f_{\rm dust}$ are small
due to the settling of dust particles toward the disk midplane. 
In the upper surface of the disk ($z>z_{\rm fric}$; see below), where the
density is low enough that the gas 
friction force does not affect the motion of dust particles, the particles 
settle in the vertical direction with the free-fall
velocity, $V_z=V_{\rm ff}=[2GM_*/(x^2+z^2)^{1/2}]^{1/2}$. In this
region the normalized dust density $\rho_{\rm dust}/\rho_{{\rm dust}, 0}$
(and the parameter $f_{\rm dust}$) drop with
decreasing $z$, inversely proportional to the gas (or the initial dust)
density. Here, the dust particles are assumed to be continuously falling on
from the surrounding molecular cloud to the disk due to the
gravitational force of the central star with a constant
(time-independent) mass flux ($=n_{\rm out}V_{\rm ff}$)
(see \S\ref{S2.1.2}). At smaller $z$ ($z<z_{\rm fric}$)
where the gas density becomes higher and
the gas friction force controls the dust motion, the vertical velocity
of dust particles becomes $V_z=\Omega_{\rm K}^2z/A\rho$
(\S\ref{S2.1.2}), and the normalized dust density $\rho_{\rm
dust}/\rho_{{\rm dust}, 0}$ (and the parameter $f_{\rm dust}$) do not
change very much in this region.  
The velocity changes from the free-fall velocity ($V_{\rm ff}$) to the
terminal velocity ($V_z=\Omega_{\rm K}^2z/A\rho$) around $z=z_{\rm
fric}=0.5$, 7.5, and 75 AU at the disk radii of $x=1$, 10, and 100AU,
respectively, in this model. 
At even smaller $z$ ($z<z_{\rm coag}$) where the density is much higher
and the  collisional cross section becomes high enough for the dust
particles to grow, the parameter $f_{\rm dust}$ drops with
decreasing $z$ (and increasing density) because small particles
disappear as a result of coagulation, while the normalized dust density
$\rho_{\rm dust}/\rho_{{\rm dust}, 0}$ does not change very much and
increases close to the disk midplane due to the settling of the
particles. Most of the dust mass settles at the disk midplane and 
$\rho_{\rm dust}/\rho_{{\rm dust}, 0}\gg 1$ at $z\approx 0$ (not shown
in this figure). Throughout the calculations the total dust mass in the
disk is equal to the initial dust mass plus the mass infalled from the
cloud (minus the mass of particles with $a>a_{\rm crit}$ removed near
the midplane of the turbulent disk; see below).
The difference between the quiescent and turbulent disks
shows up most clearly in the parameter $f_{\rm dust}$ at small $z$
($z<z_{\rm coag}$; where the dust coagulation is efficient) because
the collision rate is higher in the turbulent disk owing to the
turbulent induced relative velocity between the particles, $\delta
V_{\rm T}$ (see \S\ref{S2.1.2}). The coagulation becomes efficient
around $z=z_{\rm coag}\sim 0.15$ (0.2), 3.5 (4.0), and 65 (65) AU for
the quiescent (turbulent) disk at $x=1$, 10, and 100AU, respectively,
in this model. 

\begin{figure}
\includegraphics[scale=1.0]{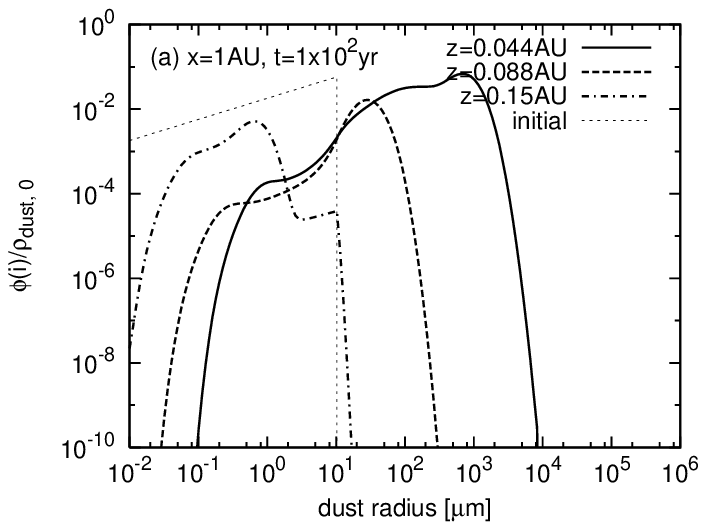}
\includegraphics[scale=1.0]{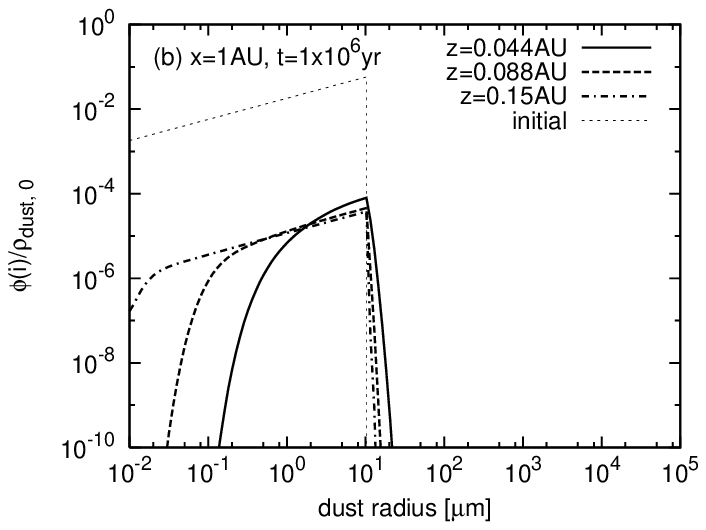}

\includegraphics[scale=1.0]{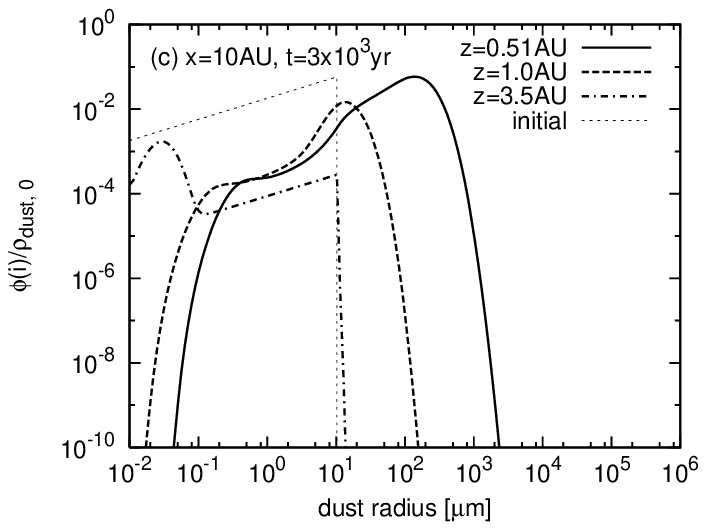}
\includegraphics[scale=1.0]{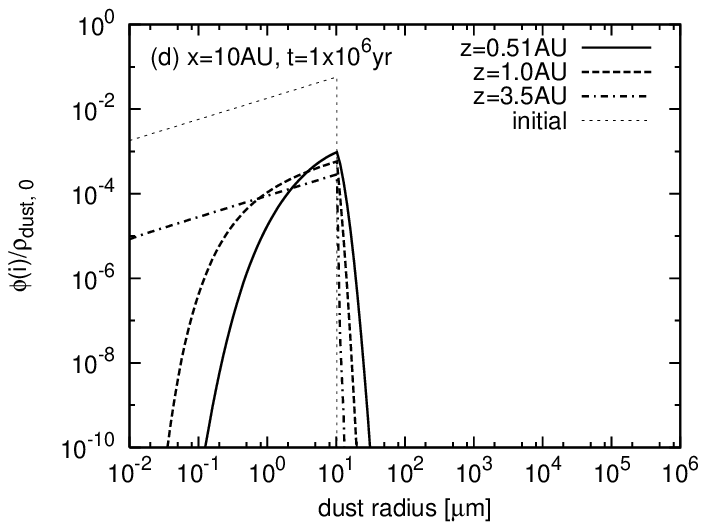}

\includegraphics[scale=1.0]{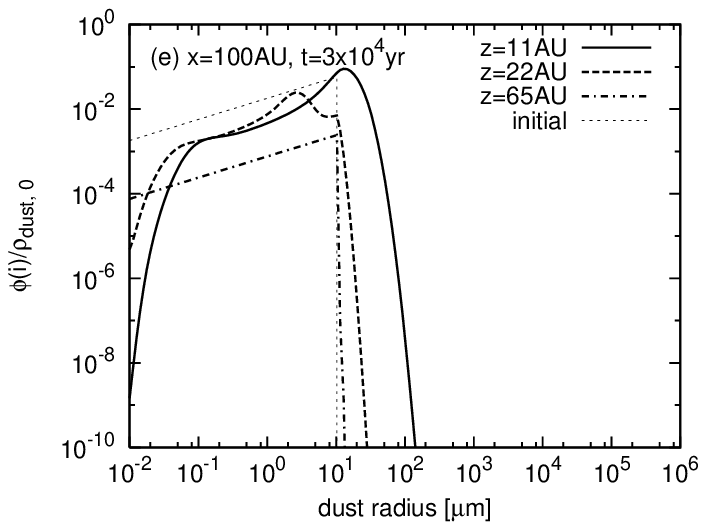}
\includegraphics[scale=1.0]{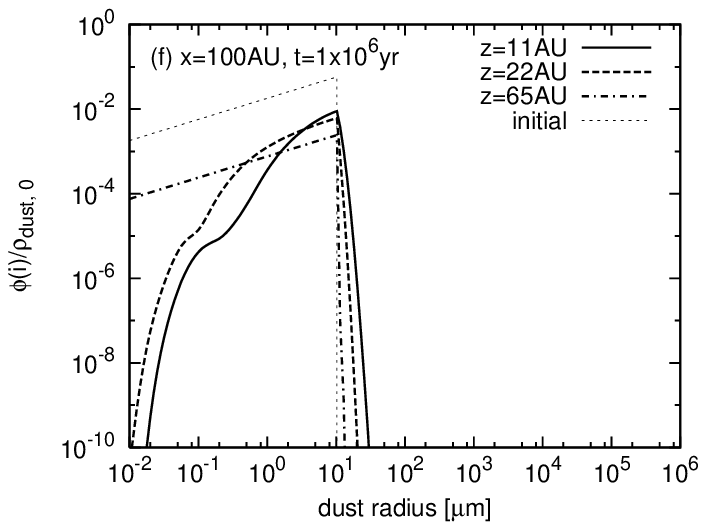}
\caption{The size distributions of mass density of dust particles,
 $\varphi(i)$, normalized by the initial dust density 
 $\rho_{\rm dust, 0}$ at each disk radii, $x$, and time, $t$, in a
 quiescent disk. The dot-dashed, dashed, and solid lines represent
 the distributions at $z\sim z_{\rm coag}$, $z\sim 2H$, and 
 $z\sim H$, respectively. The thin dotted lines show the initial
 distribution. At the disk surface the size distributions are
 similar to those in molecular clouds, but the number density of the dust
 particles is much smaller than the initial value due to the dust
 settling. Near the disk midplane small particles disappear due to the
 dust coagulation and larger particles settle towards the midplane as
 time increases. \label{f3}}
\end{figure}

\begin{figure}
\includegraphics[scale=1.0]{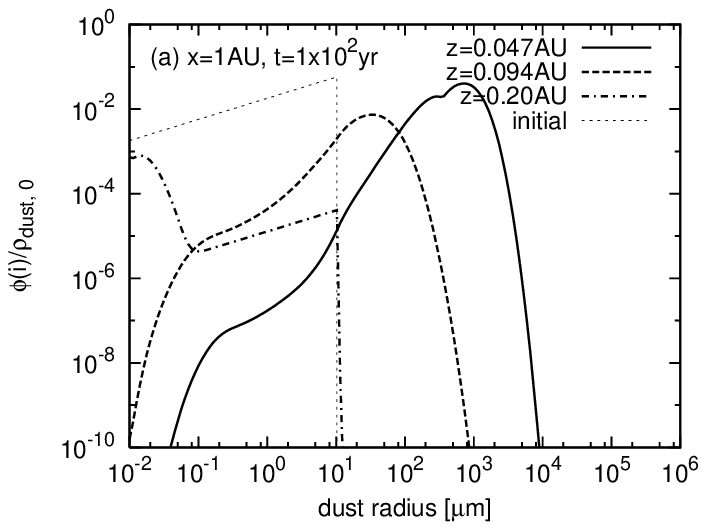}
\includegraphics[scale=1.0]{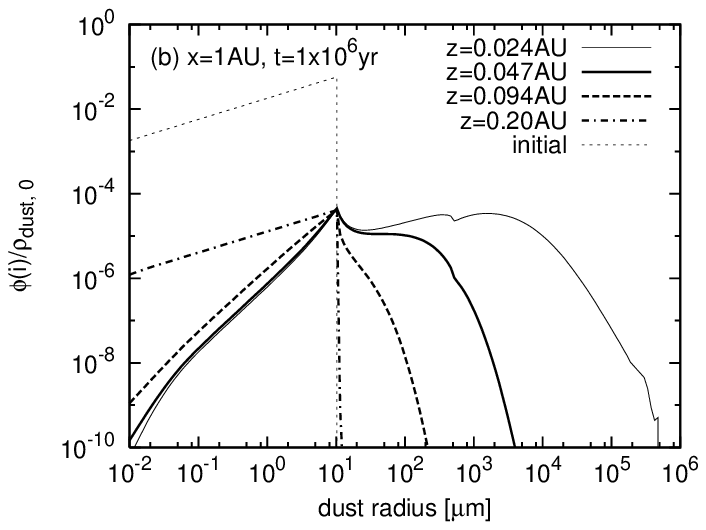}

\includegraphics[scale=1.0]{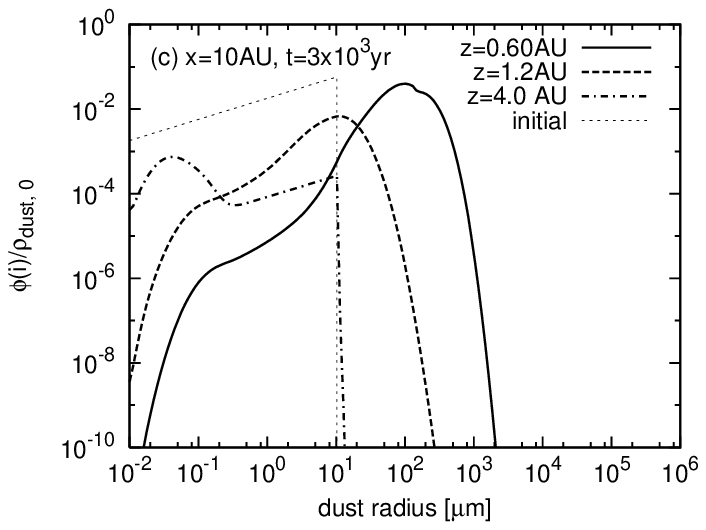}
\includegraphics[scale=1.0]{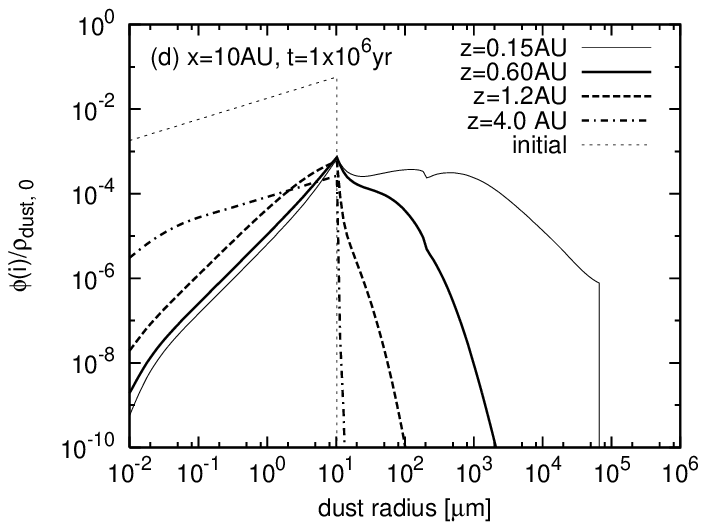}

\includegraphics[scale=1.0]{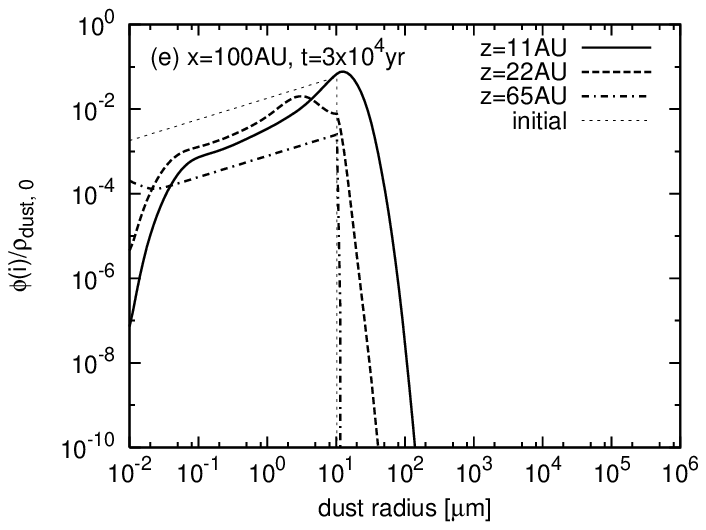}
\includegraphics[scale=1.0]{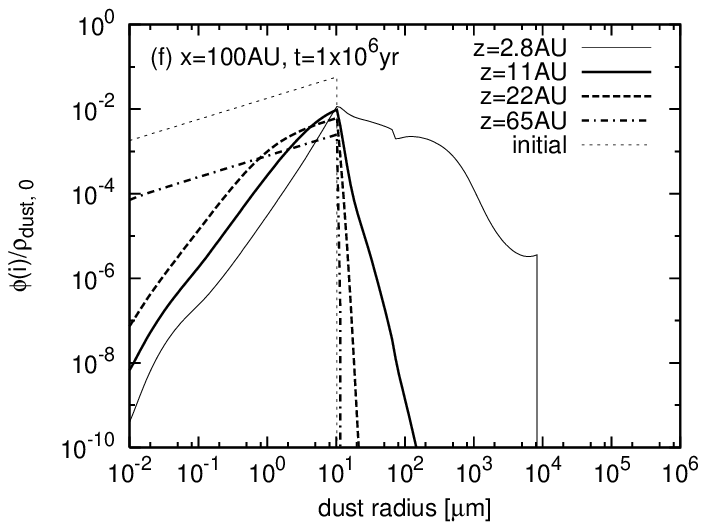}
\caption{The same as Figure \ref{f3} but in a turbulent disk. The size
distributions of dust particles at $z\sim 0.25H$ are also plotted in
 thin solid lines in Figure b, d, and f. Near the
 disk midplane a certain number of large dust particles remain due to 
 turbulent mixing. \label{f4}}
\end{figure}

In Figures \ref{f3} and \ref{f4} we plot the resulting size
distributions of mass density of dust particles, $\varphi(i)$,
normalized by the initial dust density $\rho_{{\rm dust},0}$, in
quiescent and turbulent disks, respectively. 
Each figure shows the size distributions 
at (a) $x=1$AU, $t=1\times 10^2$ yr, (b) $x=1$AU, $t=1\times 10^6$ yr; 
(c) $x=10$AU, $t=3\times 10^3$ yr, (d) $x=10$AU, $t=1\times 10^6$ yr;
(e) $x=100$AU, $t=3\times 10^4$ yr, and (f) $x=100$AU, $t=1\times 10^6$
yr. The time used in Figure {\it a, c}, and {\it e} is around
the time when the size of the largest dust particles at the disk height
of $z\sim H$ becomes maximum in the quiescent disk model.
The dot-dashed, dashed, and solid lines in each figure represent
the size distributions at $z\sim z_{\rm coag}$, $z\sim 2H$, and $z\sim
H$, respectively. In Figure \ref{f4}b, d, and f, we also plot the size
distributions at $z\sim 0.25H$ in thin solid lines.
The disk scale heights are $H=0.044$ (0.047), 0.51
(0.60), and 11 (11) AU for the quiescent (turbulent) disk at $x=1$,
10, and 100AU, respectively. The thin dotted lines show the distribution
of the initial 
%

\onecolumn

\twocolumn

\noindent
condition (the dust model A with $a_{\rm max}=10\mu$m). 
The figures show that at the surface layer
above $z\sim z_{\rm coag}$ the size distributions at $t=10^6$yr are
similar to those in dense molecular clouds (the dust model A with
$a_{\rm max}=10\mu$m in this work) as the dust particles can not
grow due to the small collisional rate, but the number density of the
particles is much smaller than the initial value due to the dust
settling as mentioned above. 
We note that bumps of mass density of small dust particles ($a\la
1\micron$) at $z\sim z_{\rm coag}$ in early phases are remnants of the
initial distribution.
At smaller $z$ ($z<z_{\rm coag}$), small dust particles disappear 
as they stick together to make larger particles. In the quiescent disk
the larger particles settle toward the disk midplane and disappear
from the disk surface, $z\geq H$, as time goes on (Fig. \ref{f3}). 
Meanwhile, in the turbulent disk a certain amount of large particles
remain even at $z\geq H$ at $t=10^6$yr (though most of them settle toward
the midplane) because of the turbulent mixing which works so as to
unify the size distributions in the vertical direction (\S\ref{S2.1.2};
Fig. \ref{f4}). 
The cutoffs around the dust radii of $a\sim$ 7 and 0.8
cm at the disk height of $z\sim 0.25H$ in 
Figure \ref{f4}{\it d} and {\it e} correspond to the critical radii,
$a_{\rm crit}$, beyond which the particles cannot be trapped in a
turbulent eddy and move toward the central star rapidly. The
particles with $a>a_{\rm crit}$ are simply removed from the calculations
(see \S\ref{S2.1.2}).


\subsection{Physical Properties of the Disks}\label{S3.2}

\begin{figure}[t]
\includegraphics[scale=1.0]{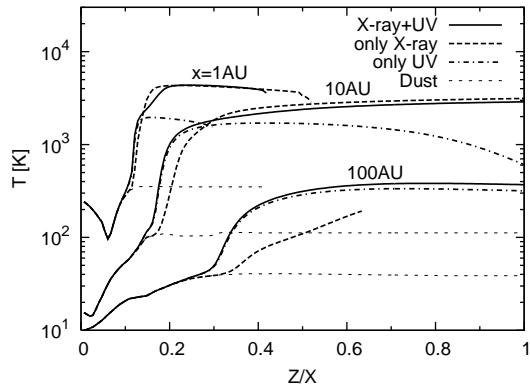}
\caption{The vertical temperature profiles of dust ({\it thin
 dotted lines}) and gas at the disk radii of $x=1, 10$ and 100AU for the
 irradiation models of X-rays $+$ UV ({\it solid lines}), X-rays only
 ({\it dashed lines}), and UV only ({\it dot-dashed lines}). The dust
 model A with $a_{\rm max}=10\micron$ is used. The X-ray
 heating is dominant at the inner region and the very surface layer of
 the disk, while the FUV heating dominates in the middle layer and the
 outer region of the disk. \label{f5}}
\end{figure}

In this subsection we obtain the gas density and temperature
distributions of the 
disk self-consistently by iteratively solving the equations for vertical
hydrostatic equilibrium and local thermal balance between heating and
cooling of gas (\S\ref{S2.3}).
The effects of the X-ray irradiation from the central star and the dust
evolution on the physical properties of the disks are discussed in the
following. 

\subsubsection{Effect of X-rays}\label{S3.2.1}

First, in Figure \ref{f5} we plot the gas temperature
profiles in the vertical direction at the disk radii of $x=1, 10$, and
100 AU, where the disk is irradiated by both of X-ray and UV radiation
from the central star ({\it solid lines}). We also plot the profiles for a
disk which is irradiated by X-ray radiation only ({\it dashed
lines}) or UV radiation only ({\it dot-dashed lines}) for comparison. 
The thin dotted lines are the dust temperature profiles which are not
affected by the UV or X-ray irradiation model. The dust model A with the
maximum dust radius of $a_{\rm 
max}=10\mu$m (see \S\ref{S2.1}) is used throughout this sub-subsection. 
We note that the calculations are performed in the region where
$\rho\geq \rho(x,z_{\infty})=5.0\times 10^{-19}$ g cm$^{-3}$, and the
position of $z_{\infty}$ depends on the models (see \S\ref{S2.3}).
The figure shows that the gas temperature is much higher than the dust
temperature in the surface layer of the disk due to the X-ray and FUV
heating. The X-ray heating dominates the FUV heating in the inner region
and in the surface layer of the disk where direct irradiation from the
central star is strong. Meanwhile,
the FUV heating dominates the X-ray heating in the middle layer and in
the outer disk. This is because the FUV radiation is scattered
efficiently by dust grains, while the Compton scattering of X-ray
radiation is inefficient in the energy range of $E\la 1$keV (e.g., Igea
\& Glassgold 1999) in which T Tauri stars mainly emit X-rays (see
Fig.~\ref{f1}). 
The gas temperature is almost the same as the dust
temperature near the disk midplane where the density is high enough 

\begin{figure}[h]
\includegraphics[scale=0.85]{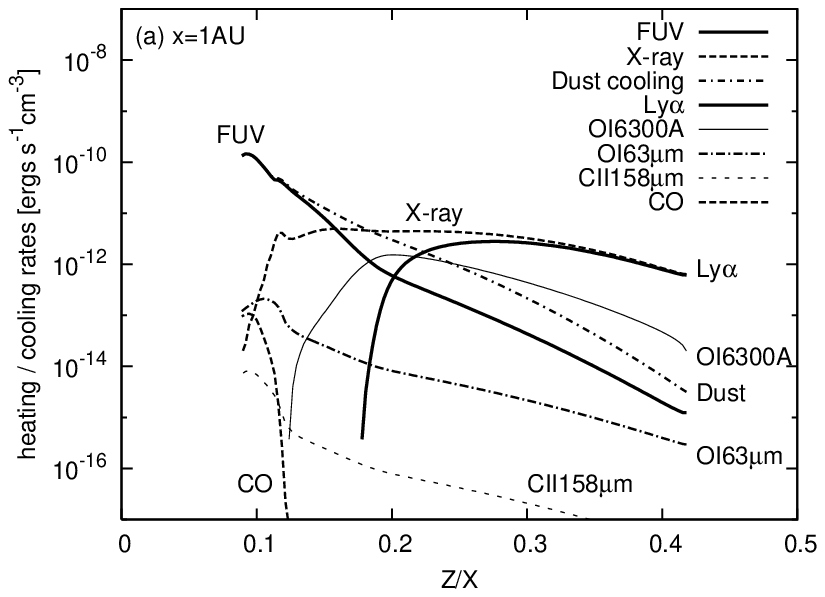}
\includegraphics[scale=0.85]{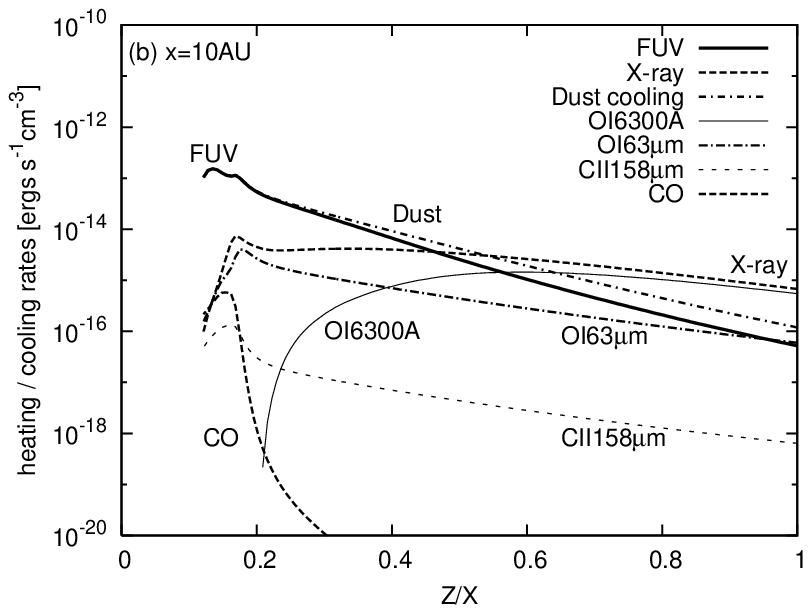}
\includegraphics[scale=0.85]{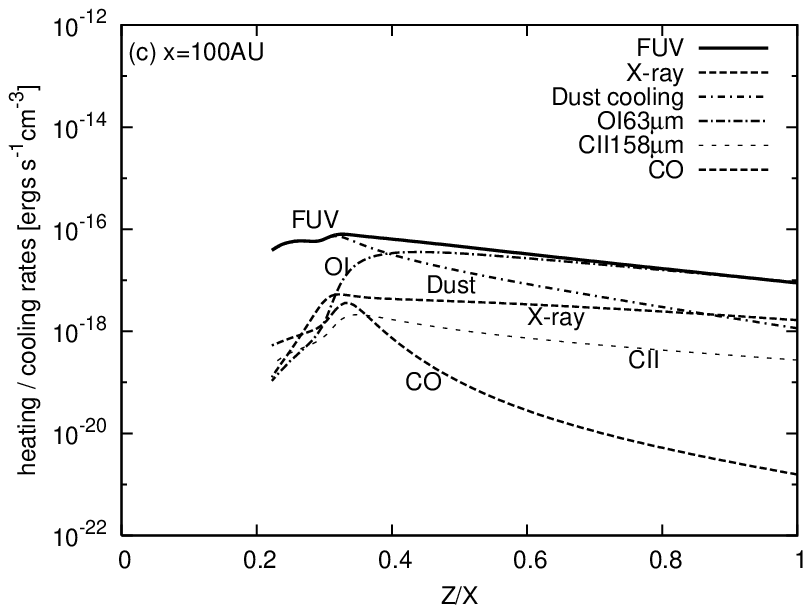}
\caption{The vertical profiles of the cooling and heating rates at the
 disk radii of (a) 
 $x=1$ AU, (b) 10 AU, and (c) 100 AU for the irradiation model of X-rays
 $+$ UV and the dust model A with $a_{\rm max}=10\micron$. The X-ray or
 FUV heating dominates the heating process, while the radiative cooling
 (Ly $\alpha$, OI 6300\AA, and OI 63$\mu$m for $x=1$, 10, and 100 AU)
 and the dust-gas collision dominate the cooling process at the surface
 layer and near the midplane, respectively. \label{f6}}
\end{figure}

\noindent
that the gas and dust particles are well coupled through collisions.

In Figure \ref{f6} we plot the vertical profiles of the heating and
cooling rates at disk radii of (a) 1AU, (b) 10AU, and (c) 100AU, for a
disk irradiated by both X-ray and UV radiation from the central
star. The figures clearly show that the X-ray heating dominates in
the inner region and in the surface layer of the disk, while the FUV
heating dominates in the middle layer and the outer region of the disk.
With regard to the cooling processes, radiative cooling dominates in 
the surface layer, while dust-gas collisions dominates near the midplane
where the density is high. The main coolant at the surface layer changes
as Ly $\alpha$, OI 6300\AA, and OI 63$\mu$m at the disk radii of 1AU,
10AU, and 100AU with decreasing gas temperature. These properties are
qualitatively the same even if we use the different dust models in
\S\ref{S2.1}. 

\begin{figure}[h]
\includegraphics[scale=1.0]{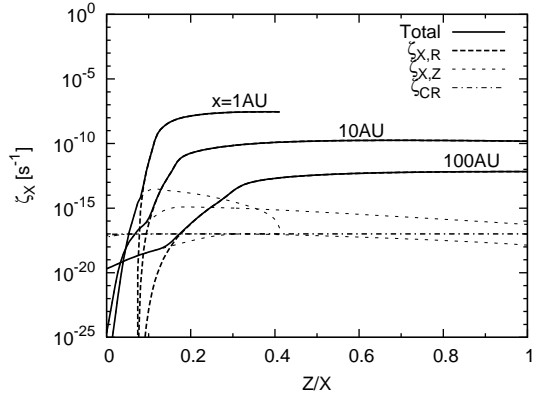}
\caption{The vertical profiles of the X-ray ionization rate at the
 disk radii of $x=1$, 10, and 100 AU for the irradiation model of X-rays
 $+$ UV and the dust model A with $a_{\rm max}=10\micron$. The solid,
 dashed, and dotted lines show the total rate, radial and vertical
 components, respectively.
 The ionization rate due to interstellar cosmic-rays are plotted as a
 dot-dashed line for comparison. The ionization rates by the X-rays from
 the central star are much lower than those by the interstellar
 cosmic-ray near the disk midplane due to the inefficient Compton
 scattering of the X-ray radiation. \label{f7}}
\end{figure}

Furthermore, we plot in Figure \ref{f7} the vertical profiles of the X-ray
ionization rates, $\zeta_{\rm X}$, defined in equation (\ref{eq.2.4.3}),
at disk radii of 1AU, 10AU, and 100AU, where the disk is irradiated
by both of X-ray and UV radiation from the central star. The radial
($\zeta_{{\rm X},R}$; {\it dashed lines}) and vertical ($\zeta_{{\rm
X},z}$; {\it dotted lines}) components, which are calculated by
substituting $F_{{\rm X},R}$ and $F_{{\rm X},z}$ of equations
(\ref{eq.2-6}) and (\ref{eq.2-7}) into equation (\ref{eq.2.4.3}) and
satisfy $\zeta_{\rm X}=\zeta_{{\rm X},R}+\zeta_{{\rm X},z}$, are also
plotted for comparison. In addition,
the ionization rates caused by interstellar cosmic-ray, $\zeta_{\rm
CR}$ are plotted in dot-dashed lines, which are estimated as 
\begin{equation}
\zeta_{\rm CR}=\zeta_{{\rm CR},0}\exp[-\Sigma(z)/\chi_{\rm CR}],
\end{equation}
where we adopt $\zeta_{{\rm CR},0}=1\times 10^{-17}$s$^{-1}$ and the
attenuation coefficient of $\chi_{\rm CR}=96$ g cm$^{-2}$ (Umebayashi \&
Nakano 1981). The surface density is calculated as
$\Sigma(z)=\int_{z}^{z_{\infty}}\rho(z')dz'$. The figure shows that at
the disk surface 
the ionization rates due to X-rays from the central star 
are much higher than those due to interstellar cosmic-rays, while
near the disk midplane the former is much lower than the latter. This is
because X-ray attenuation is larger than that of cosmic-rays and because
the Compton scattering of X-ray radiation is inefficient (see
Fig.~\ref{f1} and Igea \& Glassgold 1999). 

\clearpage


\subsubsection{Effect of Dust Evolution}\label{S3.2.2}

\begin{figure}
\includegraphics[scale=1.0]{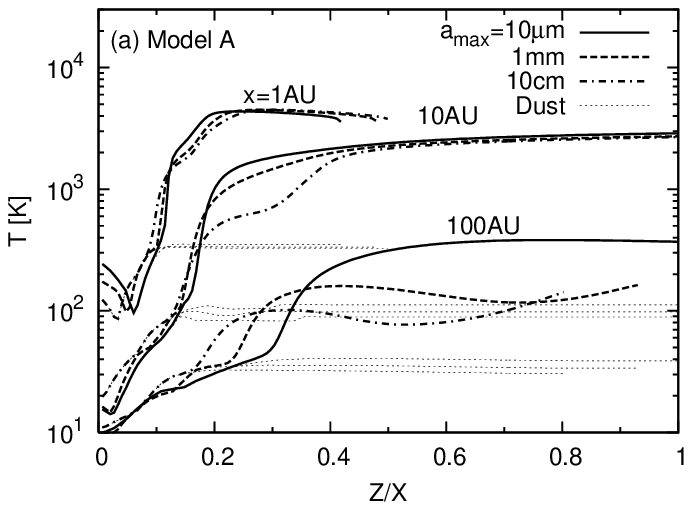}
\includegraphics[scale=1.0]{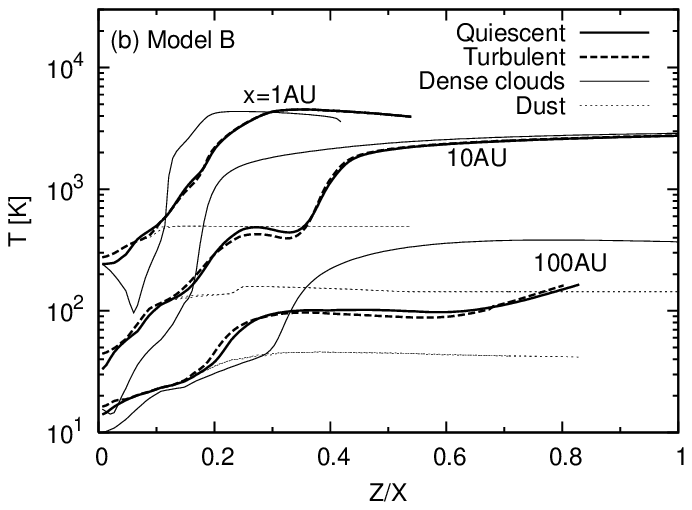}
\caption{The vertical temperature profiles of dust ({\it thin dotted
 lines}) and gas at the disk radii of $x=1, 10$ and 100AU for (a) 
 dust model  
 A with $a_{\rm max}=10\micron$ ({\it solid lines}), 1mm ({\it dashed
 lines}), and 10cm ({\it dot-dashed lines}), and (b) model B in
 quiescent ({\it solid lines}) and turbulent ({\it dashed lines}) disks.
 The profiles for the dense cloud dust model (dust
 model A with $a_{\rm max}=10\micron$) are plotted here as
 thin solid lines. The irradiation model of X-rays $+$ UV is used here.
 As the dust particles grow or settle toward the disk midplane 
 the gas temperature in the middle layer and in the outer disk decreases 
 due to the decrease of the grain photoelectric
 heating rate. Meanwhile, the dust and gas temperatures near the midplane
 increases due to smaller grain opacity and greater penetration
 of the irradiation from the central star. \label{f8}}
\end{figure}

In Figure \ref{f8} we plot the vertical gas temperature profiles for 
various dust models at disk radii of $x=1, 10$, and 100
AU for the case of a disk heated by both X-rays and UV radiation. 
The profiles for dust model A with different maximum dust radii
of $a_{\rm max}=10\mu$m ({\it solid lines}), 1mm ({\it dashed lines}),
and 10cm ({\it dot-dashed lines}) are plotted together in
Figure~\ref{f8}a. The profiles for dust model B at $10^6$yr after
the calculation starts are plotted in Figure~\ref{f8}b for the quiescent
({\it solid lines}) and turbulent ({\it dot-dashed lines}) disks. 
The profiles calculated by using the dense cloud dust model (which is
the initial condition of the calculation for the dust evolution and the
dust model A with $a_{\rm max}=10\mu$m) are also plotted together in
thin solid lines for comparison. The thin dotted 
lines in the figures are the dust temperature profiles. 
Figure~\ref{f8}a shows that as the dust
particles grow and the total surface area of dust grains ($f_{\rm
dust}$) decreases (see \S\ref{S2.1.3}), the gas temperature at the disk 
surface drops because the grain photoelectric heating rate decreases
(e.g., Aikawa \& Nomura 2006). The gas temperatures in the
inner disk ($x\sim 1$ AU) and in the surface layer at $x\sim 10$ AU
do not change because the X-ray heating dominates in these regions.
The dust temperature at the disk surface decreases slightly with
dust growth.
Figure \ref{f8}b shows that clear differences appear in the gas and dust
temperatures between the dust model of dense clouds (dust model A
with $a_{\rm max}=10\mu$m) and the models with the dust evolution in
both quiescent and turbulent disks. For the models with dust
evolution the gas temperature in the middle
layer and the outer region of the disks drops owing to the decrease of
$f_{\rm dust}$ (see \S\ref{S2.1.3} and \S\ref{S3.1}), while the dust and
gas temperatures near the midplane increase
because the grain opacity decreases and the irradiation from
the central star can penetrate deeper in to the disks. At $x\sim 1$AU 
heating via irradiation dominates even near the midplane for the
models with the dust evolution, whereas the viscous heating is dominant for
the dense cloud dust model. The differences between the quiescent and
turbulent disks are small because the profiles of $f_{\rm dust}$ are
similar, especially in the surface layer. The dust and gas temperatures
very close to the midplane are slightly higher in the turbulent disk
due to the higher collision rate between the dust particles, which results
in lower $f_{\rm dust}$ and grain opacity (see \S\ref{S3.1}). 

Dust growth and settling are also expected to impact on the
X-ray heating rates and the gas temperature profile through the
change in the photoionization cross section, $\sigma_{\rm ph}$, part of
which is contributed by heavy elements in dust
grains (e.g., Glassgold et al. 1997; Wilms et al. 2000). Here we check
the effects by simply adopting an extreme case, specifically that the 
contribution of dust grains to the cross section is negligible
for dust model A with $a_{\rm max}=10$cm and for dust model B.
We modify the cross section in Malony et al. (1996) 
by simply assuming that the contribution by heavy elements in gas phase
is about 60\%, on average, of the total cross section 
(Wilms et al. 2000). 
This modification of the cross section makes the gas temperature 
higher or lower by a factor of 2 at most
at the disk surface, where the X-ray heating is dominant. 
At large $z$ the gas temperature becomes
slightly lower due to the decrease of the photoionization rate, while at
smaller $z$, where the influence of attenuation is more important, the
temperature becomes a bit higher owing to the decrease of the attenuation
coefficient, which results in the relatively 
stronger X-ray radiation field (e.g., Glassgold et al. 1997).
The gas density at the disk surface also becomes higher or lower by a
factor of 3 at most,
according to the change in the gas temperature.
The variation of the photoionization cross section due to the dust
evolution will also slightly affect the level populations and the line
emission of molecular hydrogen through the changes in the thermal
collision and the X-ray pumping rates. When the FUV heating or the UV
pumping process dominates, however, the changes will be small.
In the following sections we neglect these effects for simplicity. 

\begin{figure}[t]
\includegraphics[scale=1.0]{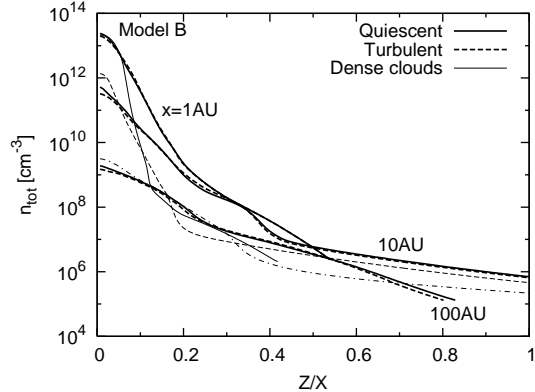}
\caption{The vertical gas density profiles at the disk radii of 
 $x=1, 10$ and 100AU for dust model B in quiescent ({\it solid
 lines}) and turbulent ({\it dashed lines}) disks. The profiles for the
 dense cloud dust model are also plotted together in thin solid 
 ($x=1$AU), dotted ($x=10$AU), and dot-dashed ($x=100$AU) lines.
 The disks are puffed out more for the models with the dust evolution
 due to higher gas temperatures at the disk midplane and
 higher disk scale height. \label{f9}}
\end{figure}

As the dust particles evolve in the disk, the gas density profile also
changes since it is related to the gas temperature
profile. In Figure \ref{f9} we plot the gas density profiles in the
vertical direction at the disk radii of $x=1, 10$, and 100AU, which are
calculated by
using dust model B in quiescent ({\it solid lines}) and turbulent
({\it dashed lines}) disks. The profiles for the dense cloud dust model 
are also plotted together as thin solid ($x=1$AU), dotted 
($x=10$AU), and dot-dashed ($x=100$AU) lines for comparison. The gas
densities in the models 

\onecolumn

\begin{figure}
\includegraphics[scale=0.44]{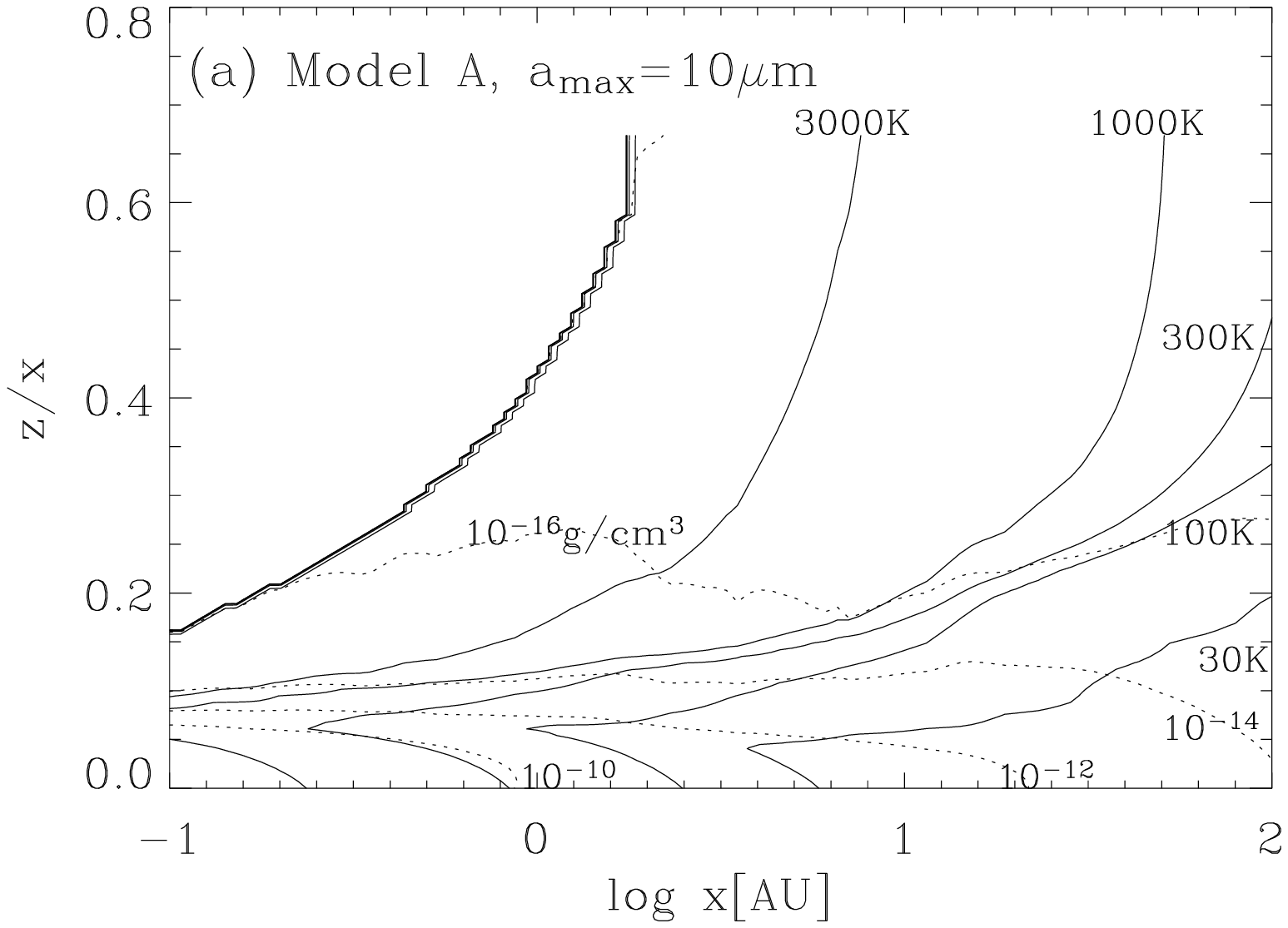}
\includegraphics[scale=0.44]{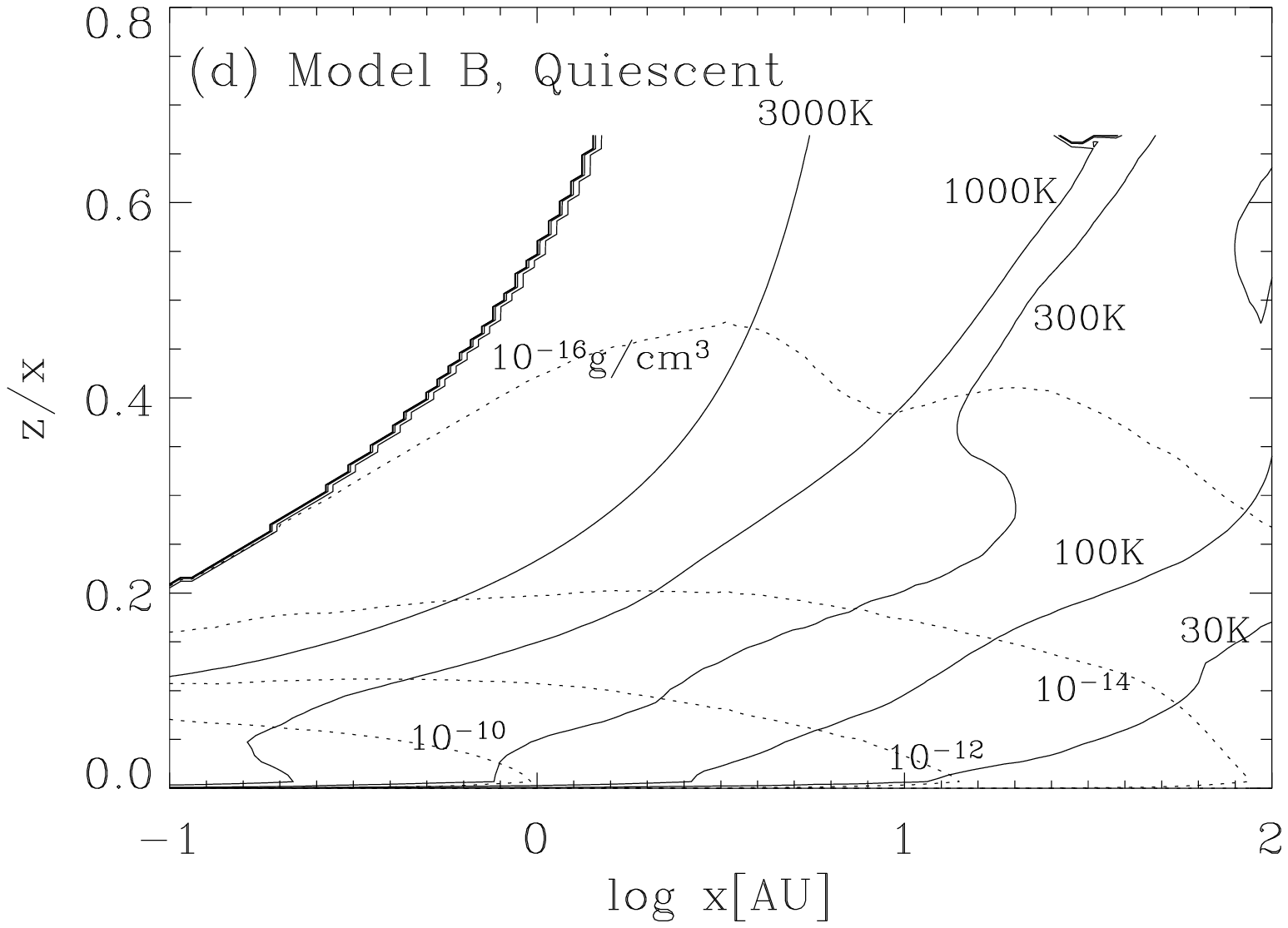}

\includegraphics[scale=0.44]{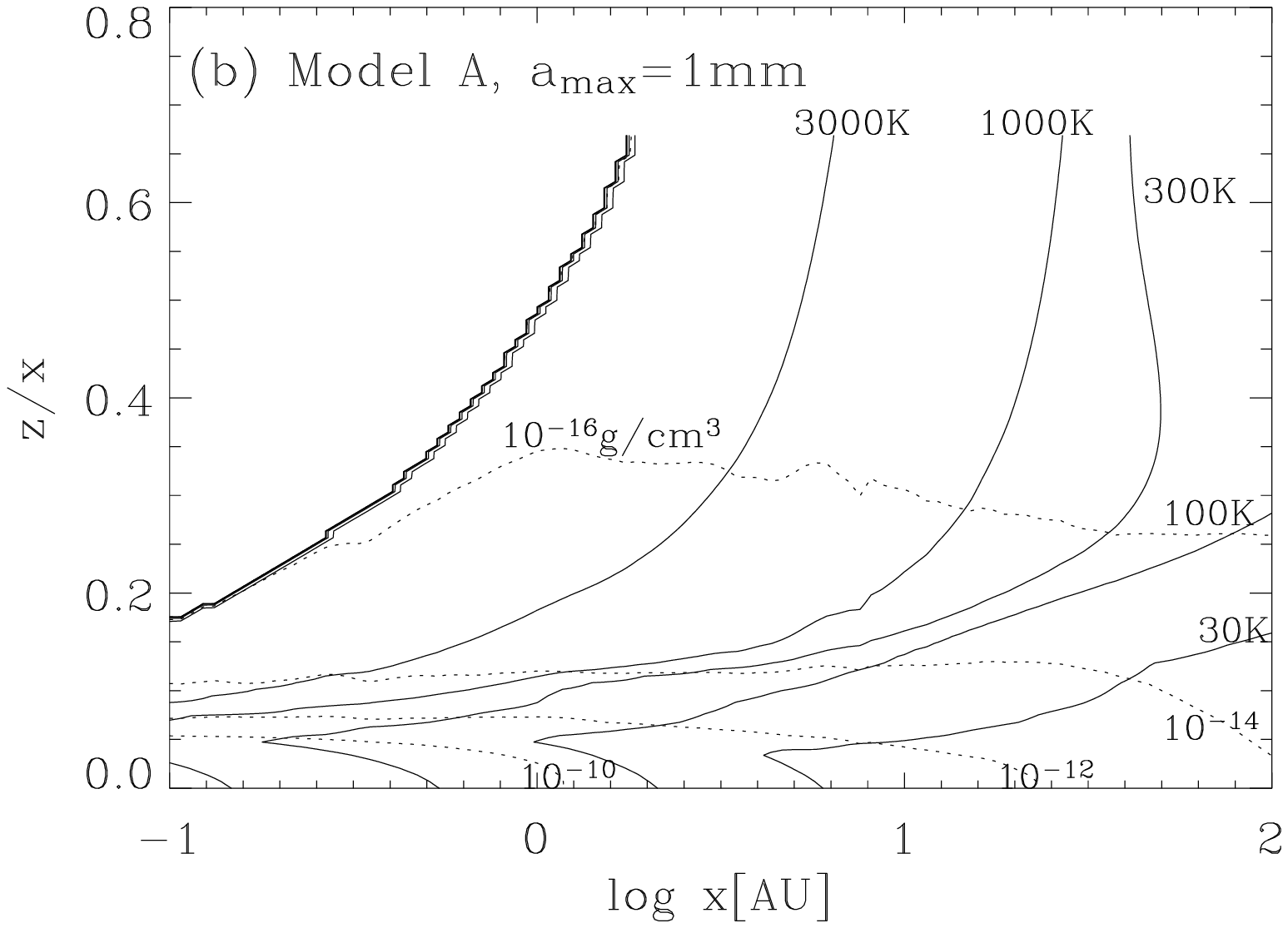}
\includegraphics[scale=0.44]{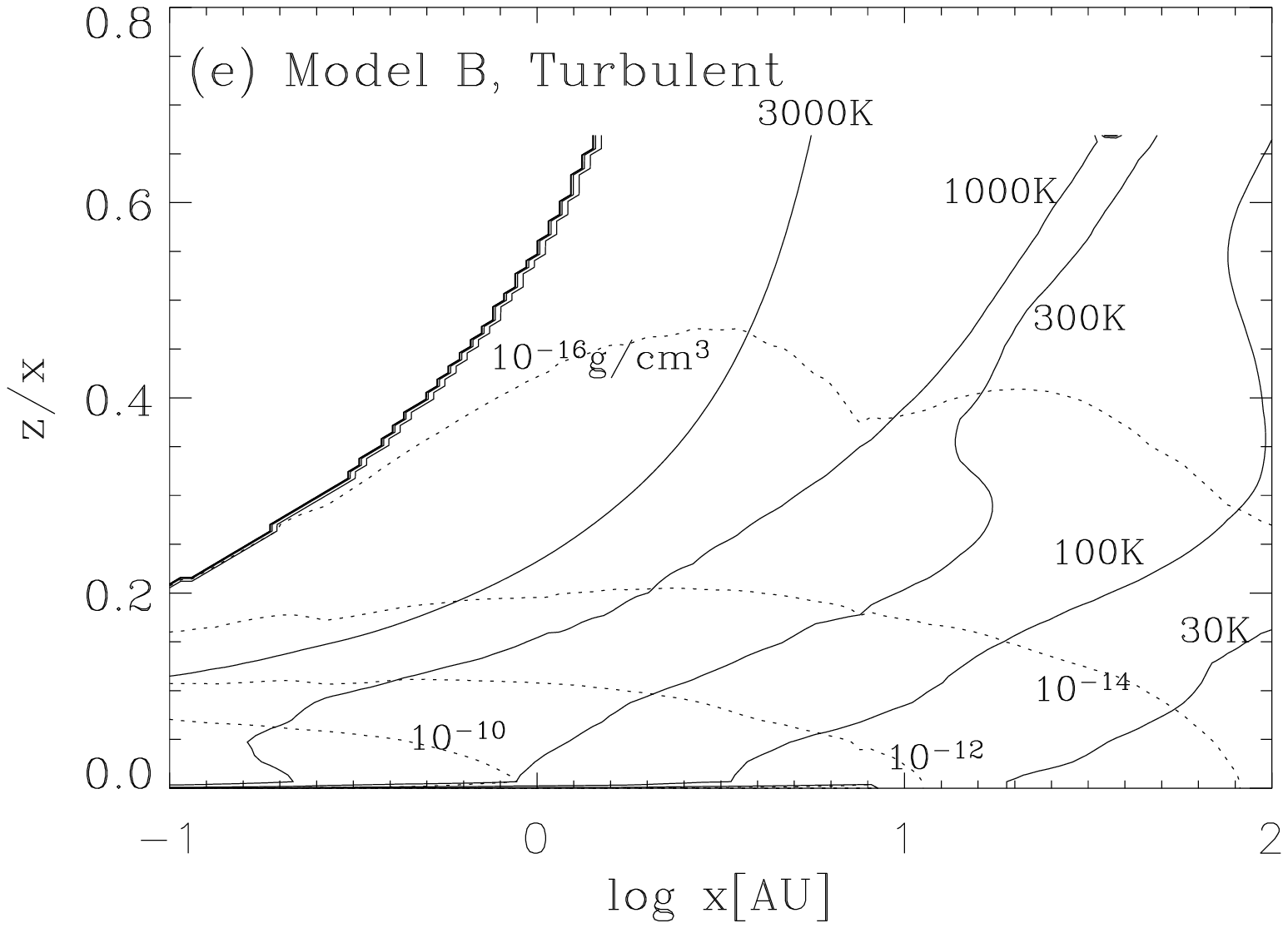}

\includegraphics[scale=0.44]{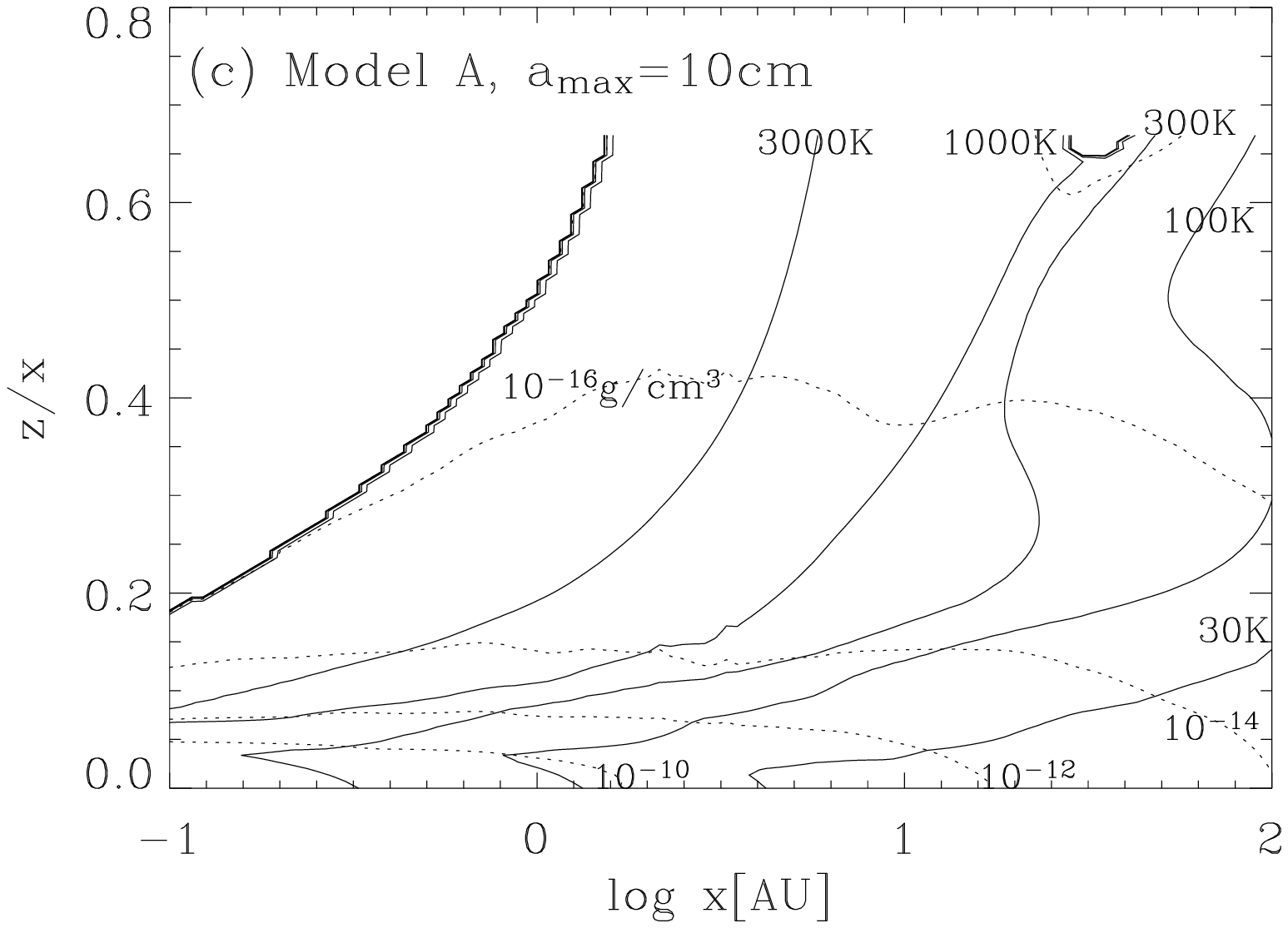}
\caption{The contour plots of the gas temperature ({\it solid lines})
 and density ({\it dotted lines}) distributions in the $z/x$ vs. $z$
 plane for dust model A with (a) $a_{\rm max}=10\micron$, (b) 1mm,
 and (c) 10cm, and model B in (d) quiescent and (e) turbulent
 disks. The irradiation model of X-rays $+$ UV is used here. \label{f10}}
\end{figure}

\twocolumn


\noindent
with the dust evolution are lower at the disk
midplane and higher at the disk surface than those for the dense cloud
dust model because of higher gas temperatures at the midplane and
higher disk scale height ($H=\cs_0/\Omega_{\rm K}$). 
In Figure \ref{f10} we present the contour plots of the resulting gas
temperature ({\it solid lines}) and density ({\it dashed lines})
profiles in the $z/x$ vs. $x$ plane. The contour levels are taken as
$T=30,100,300,1000$, and $3000$K, 
and $\rho=10^{-16},10^{-14},10^{-12}$, and $10^{-10}$g cm$^{-3}$. 
Dust model A with (a) $a_{\rm max}=10\mu$m, (b) 1mm, and
(c) 10cm, and model B in (d) quiescent and (e) turbulent
disks are used in these figures. 

Making use of the obtained density and temperature distributions, we
calculate the continuum radiation of thermal dust emission from the
disks by solving the radiative transfer equation, simply assuming that
the disks are face on to an observer. The resulting infrared (IR) spectra
basically reproduce the median spectral energy
distribution (SED) observed toward classical T Tauri stars (CTTSs) in
the Taurus-Auriga molecular cloud (D'Alessio et al. 1999, 2006) 
for the dust model A with $a_{\rm
max}=10\micron$. For the models with larger maximum dust radii, the
resulting IR dust emission is weaker than the median SED
by a factor of about 4 at the most (e.g., D'Alessio et al. 2001).
The thermal dust emission from the disks for dust model B in both
quiescent and turbulent disks also reproduces the median observed
SED if we adjust the inner disk radii. These models do not reproduce the 
flux deficits relative to the median SED in the near-IR to the mid-IR
wavelength bands which are observed toward several CTTSs, including
TW Hya (e.g., Calvet et al. 2002; Bergin et al. 2004). However, the disk
structure beyond several tens of AU where molecular hydrogen lines are
mainly emitted (see Paper I) will be almost unaffected even if
we were to modify the structure of inner disk within several AU in order
to force the thermal dust emission to account for the flux deficits,
because only a limited region close to the midplane of the outer
disk can be shadowed by the inner disk since the disk has a flared
structure. 

\begin{figure}[t]
\includegraphics[scale=1.0]{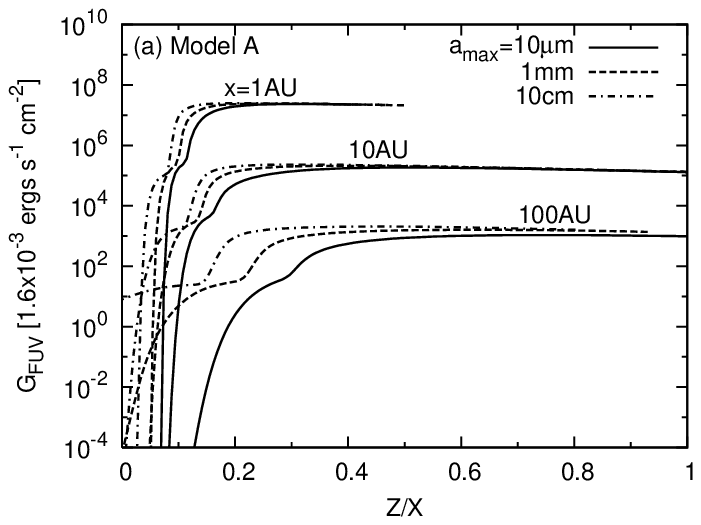}
\includegraphics[scale=1.0]{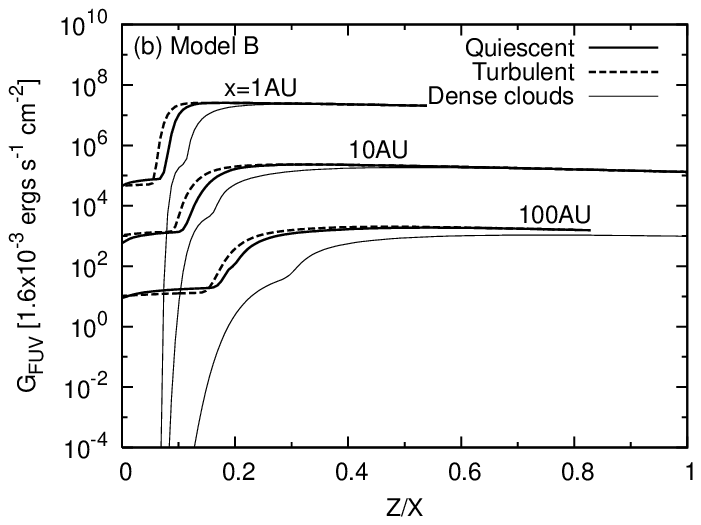}
\caption{The vertical profiles of the integrated FUV radiation fields
 (6eV $<h\nu <$ 13eV) at the disk radii of $x=1, 10$ and 100AU for
 dust models A (a) and B (b). The profiles are plotted in the same way as
 in Figure \ref{f8}. The FUV radiation from the central star
 penetrates deeper in the disks as the dust particles evolve and the
 grain opacity decreases. \label{f11}}
\end{figure}

Finally, in Figure \ref{f11} we show
another effect of the dust evolution due to the change in grain opacity.
The figure shows the profiles of the integrated FUV radiation fields for
the energy range of 6eV $<h\nu <$ 13eV 
in the vertical
direction at the disk radii of $x=1, 10$, and 100 AU.
The profiles for dust models A and B are plotted in Figure
\ref{f11}a and b, respectively, in the same way as Figure \ref{f8}.
The figures show that as the dust particles evolve in the disk and
$f_{\rm dust}$ decreases, the FUV radiation from the central star
penetrates deeper in the disk due to the decrease of grain opacity
(see also Paper I for the FUV radiation fields in disks).

\subsection{Level Populations of Molecular Hydrogen}\label{S3.3}

Making use of the physical properties of the disks obtained in the
previous subsections, we calculate the level populations of molecular
hydrogen in the disks by solving the equations for statistical
equilibrium (\S\ref{S2.4}). The effects of X-ray irradiation and dust 
evolution on the level populations are discussed in the following.

\begin{figure}[h]
\includegraphics[scale=1.0]{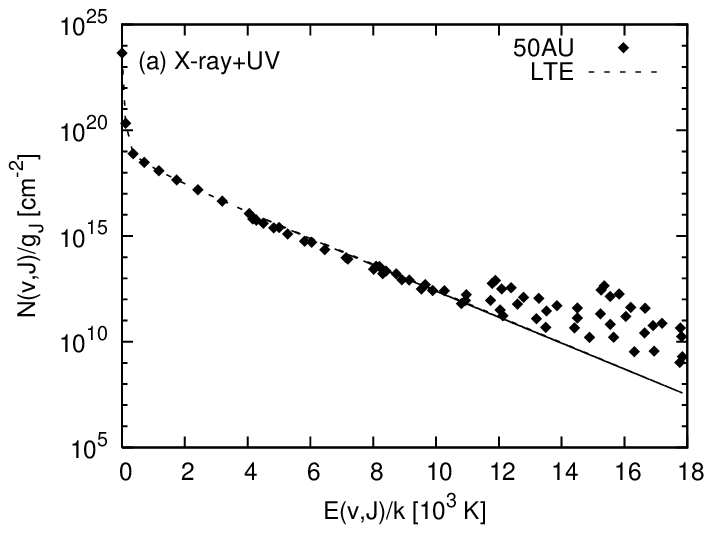}
\includegraphics[scale=1.0]{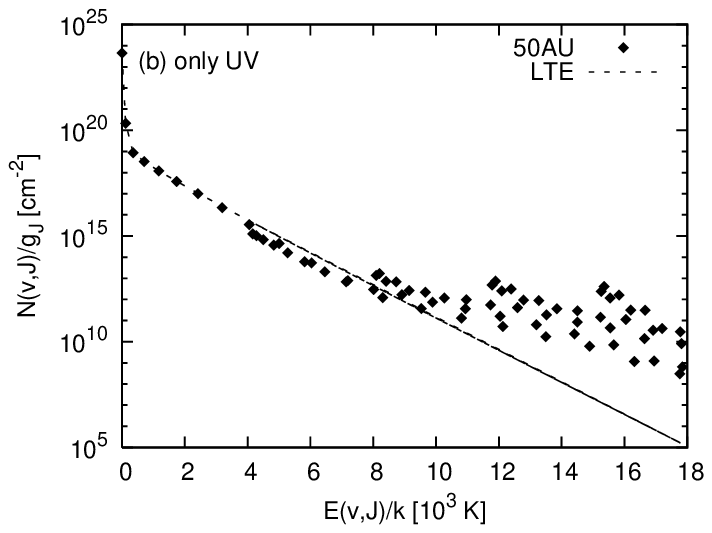}
\includegraphics[scale=1.0]{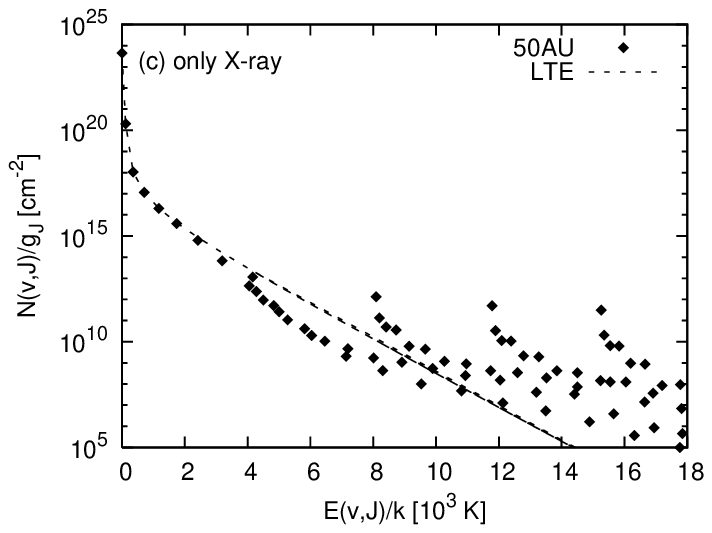}
\caption{The level populations of molecular hydrogen at a disk radius
 of 50 AU ({\it filled diamonds}) for the irradiation models of (a)
 X-rays $+$ UV, (b) UV only, and (c) X-rays only. The LTE distributions
 are plotted as dashed lines. Dust model A with $a_{\rm max}=10\micron$
 is used here. The populations are in LTE in lower energy
 levels when the disk is irradiated by strong UV radiation, 
while they are controlled by X-ray pumping if the UV 
 irradiation is weak and X-ray irradiation is strong. \label{f12}}
\end{figure}

\subsubsection{Effect of X-rays}

In Figure \ref{f12} we plot the resulting level populations of
molecular hydrogen for the models in which the disk is irradiated by
(a) both X-ray and UV radiation from the central star, (b) UV
radiation only, and (c) X-ray radiation only. Dust model A with 
maximum dust radius of $a_{\rm max}=10\mu$m (see \S\ref{S2.1}) is used in
this sub-subsection. 
The filled diamonds show the column densities of molecular hydrogen in
each ro-vibrational level as a function of the level energy.
The column densities are calculated by integrating the number density of
molecular hydrogen in each level along the vertical direction at a
disk radius of 50AU. The level populations in local thermodynamic
equilibrium (LTE) are shown as dashed lines. 
The figure shows that if we take into account UV irradiation from
the central star, the gas temperature becomes high enough for the
collisional excitation process to be very
efficient, and the level populations in lower energy levels are in LTE
distribution as a result. Meanwhile, if the disk is irradiated by
X-rays only and the gas is cold enough, the populations are not in LTE
due to the X-ray pumping process. It suggests that we may
be able to observe molecular hydrogen transitions excited by
X-ray pumping toward those protoplanetary disks whose central stars have
strong X-ray and weak UV radiation.

\subsubsection{Effect of Dust Evolution}\label{S3.3.2}

\begin{figure}
\includegraphics[scale=1.0]{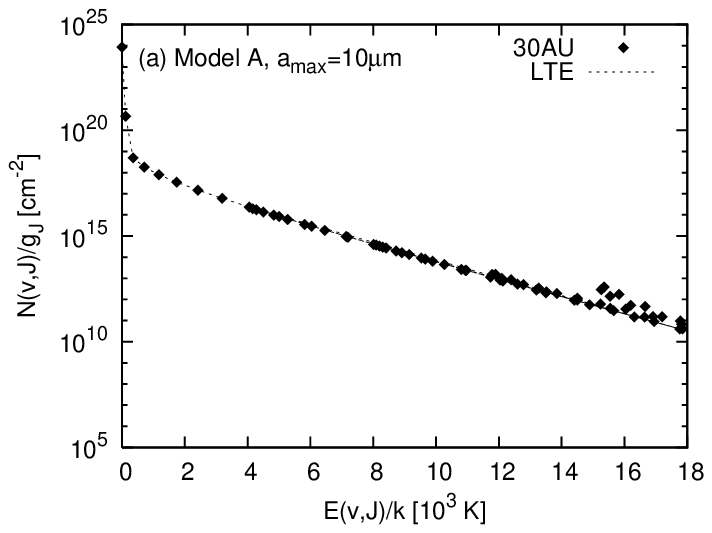}
\includegraphics[scale=1.0]{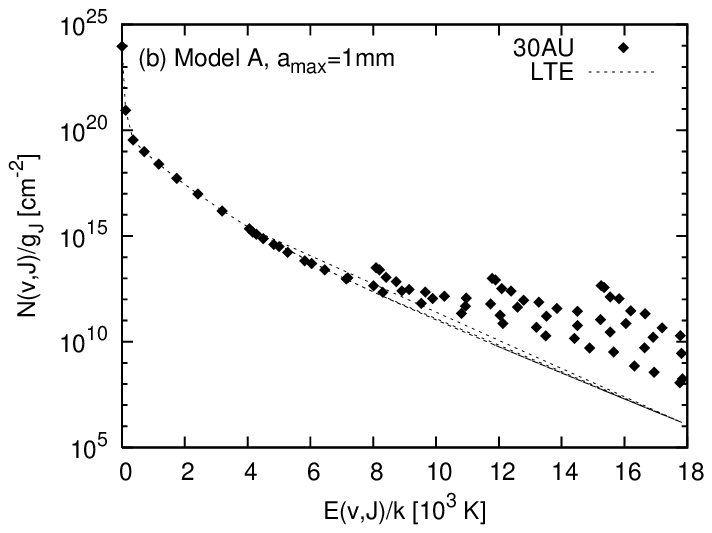}

\includegraphics[scale=1.0]{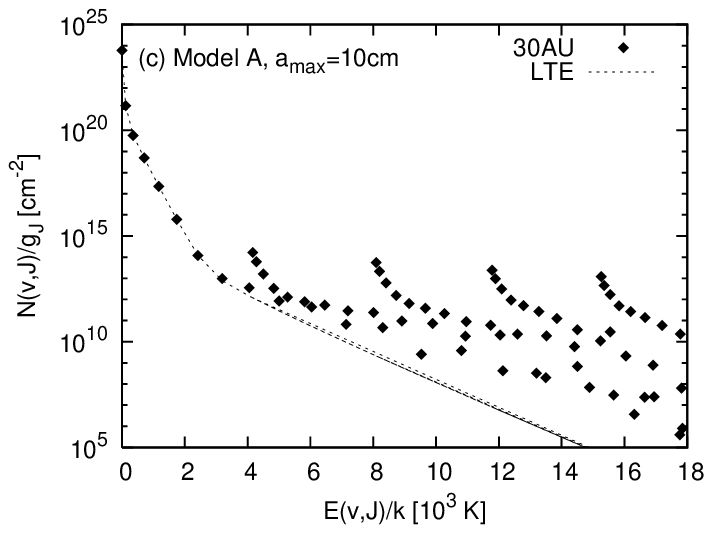}
\includegraphics[scale=1.0]{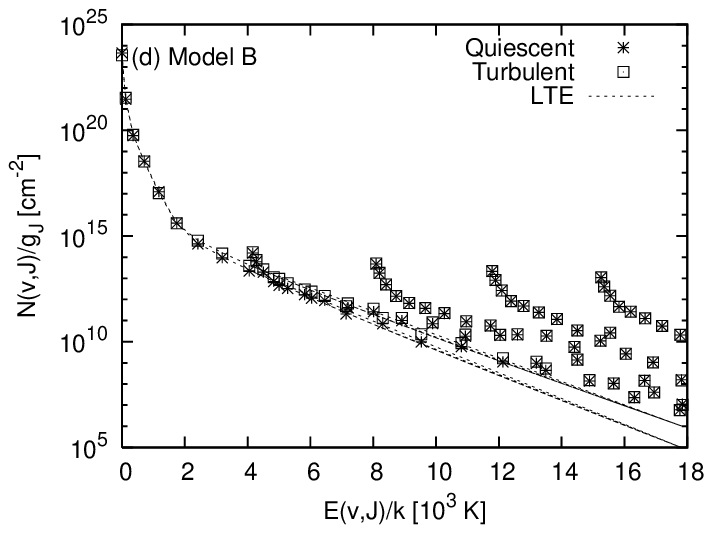}
\caption{The level populations of molecular hydrogen at a disk radius
 of 30 AU for dust model A ({\it filled diamonds}) with 
 $a_{\rm max}=$ (a) 10$\micron$, (b) 1mm, and (c) 10cm, and (d)
 model B in 
 quiescent ({\it asterisks}) and turbulent ({\it open squares}) disks.
 The irradiation model of X-rays $+$ UV is used here.
 The level populations change from LTE to non-LTE as
 dust particles grow or settle toward the disk midplane, and since the
 gas temperature drops while the UV photons in the disk increase.
 \label{f13}}
\end{figure}

We now discuss the effect of dust evolution on the level
populations of molecular hydrogen in a disk irradiated by both X-rays 
and UV radiation. Figure \ref{f13} is the same as
Figure \ref{f12} but for different dust models. Dust model A
with maximum dust radii of $a_{\rm max}=$ (a) 10$\micron$, (b) 1mm,
and (c) 10cm and (d) model B are used in these figures. 
Figures \ref{f13}a-c show that as the dust particles grow, the level
populations of molecular hydrogen change from LTE to non-LTE
distributions. This is because with increasing dust size and decreasing
$f_{\rm dust}$, the gas temperature drops due to the decrease of grain
photoelectric heating rate, and the collisional excitation
process becomes less efficient. In addition, the UV radiation from the

\onecolumn

\twocolumn

\noindent
central star can penetrate deeper in the disk due to the decrease of
grain opacity, and the UV pumping process becomes more efficient.
X-ray pumping is not the dominant process if we take into account
the UV irradiation from the central star.
In Figures \ref{f13}d the level populations with dust model B in
quiescent ({\it asterisks}) and turbulent ({\it open squares}) disks are
plotted together. The populations in quiescent and turbulent disks are
almost identical because of similar physical properties of the disks
(see \S\ref{S3.2.2}). The figure shows that if we take into account
dust evolution,
the level populations are in non-LTE distributions both in quiescent and
turbulent disks because of low $f_{\rm dust}$, which results in low gas
temperature and high UV radiation fields, while they are in LTE
distributions with the dust model of dense clouds (dust model A with
$a_{\rm max}=10\mu$m).

\subsection{Molecular Hydrogen Emission}\label{S3.4}

Making use of the physical properties of the disks and the level
populations of molecular hydrogen obtained in the previous
subsections, we calculate the line emission from molecular hydrogen
(\S\ref{S2.4}). In the following we show the resulting line
spectra in the near- and mid-infrared (NIR and MIR), and ultraviolet
(UV) wavelength bands, and present line ratios, using various dust and
irradiation models.

\subsubsection{Line Spectra}

\begin{figure}
\includegraphics[scale=1.0]{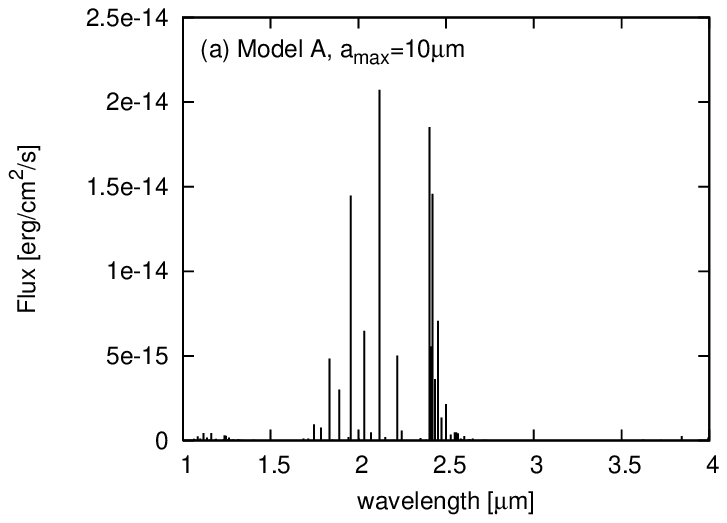}
\includegraphics[scale=1.0]{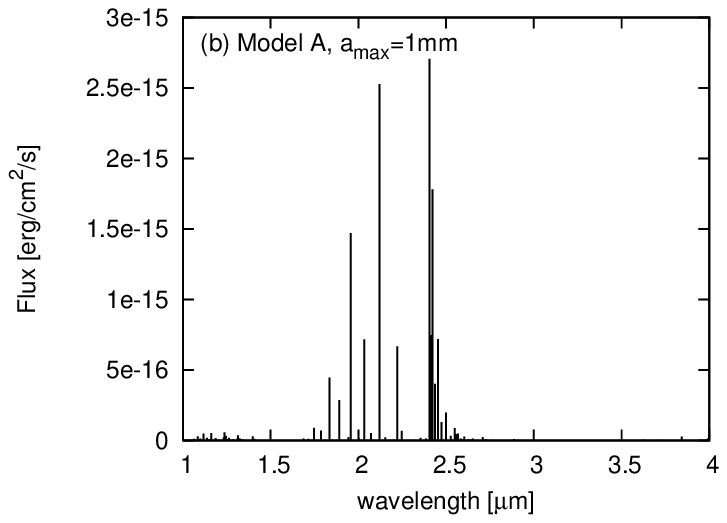}

\includegraphics[scale=1.0]{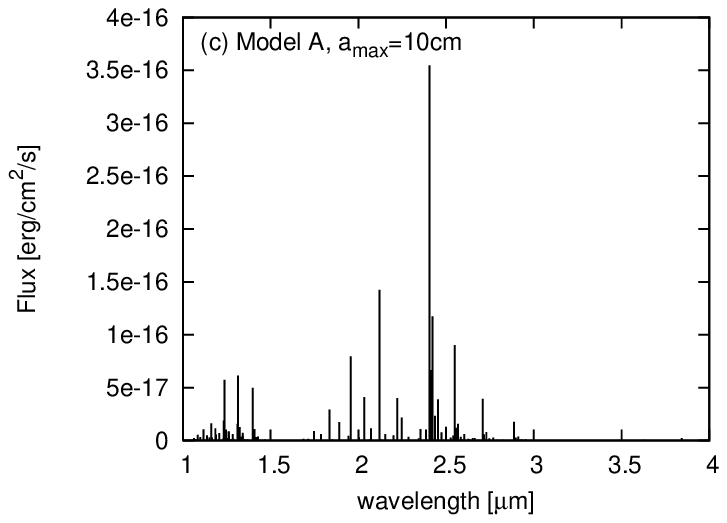}
\includegraphics[scale=1.0]{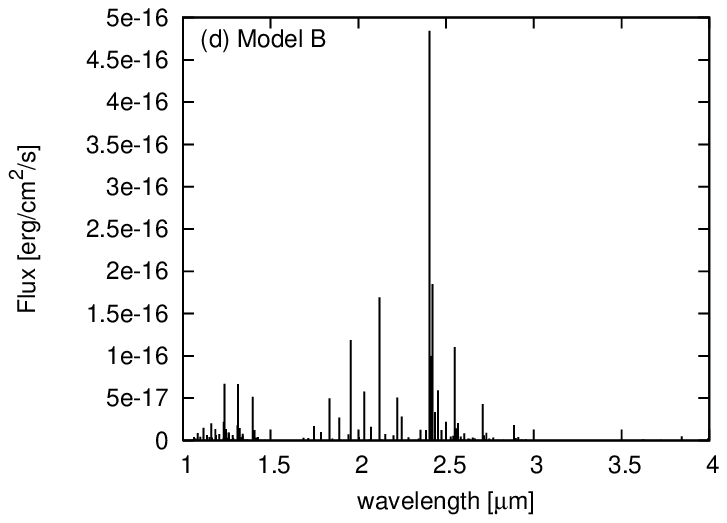}
\caption{The near-infrared ($1\micron <\lambda <4\micron$) spectra of
 ro-vibrational transition lines of molecular hydrogen from the disks for
 dust model A with $a_{\rm max}=$ (a) $10\micron$, (b) 1mm, and (c)
 10cm, and (d) model B in a quiescent disk. 
 The irradiation model of X-rays $+$ UV is used here.
 The distance to the disk is set to be $d=56$pc.
 The lines from higher
 energy levels become relatively stronger as the dust particles evolve
 and the level populations change from LTE to non-LTE. \label{f14}}
\end{figure}

\begin{figure}
\includegraphics[scale=1.0]{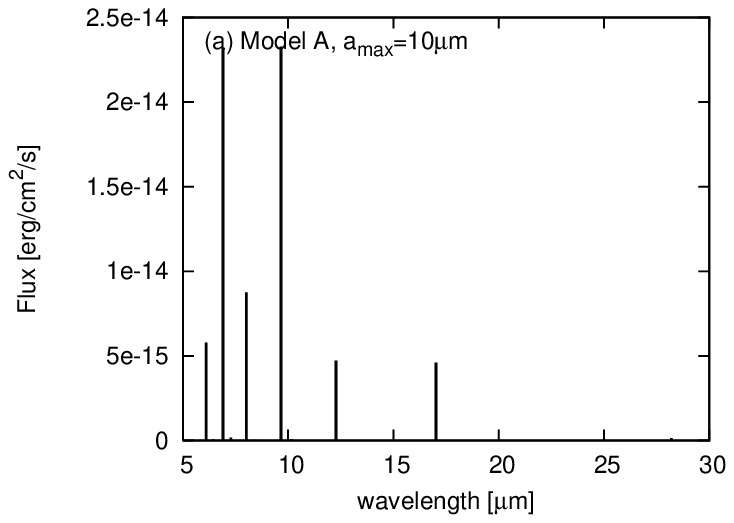}
\includegraphics[scale=1.0]{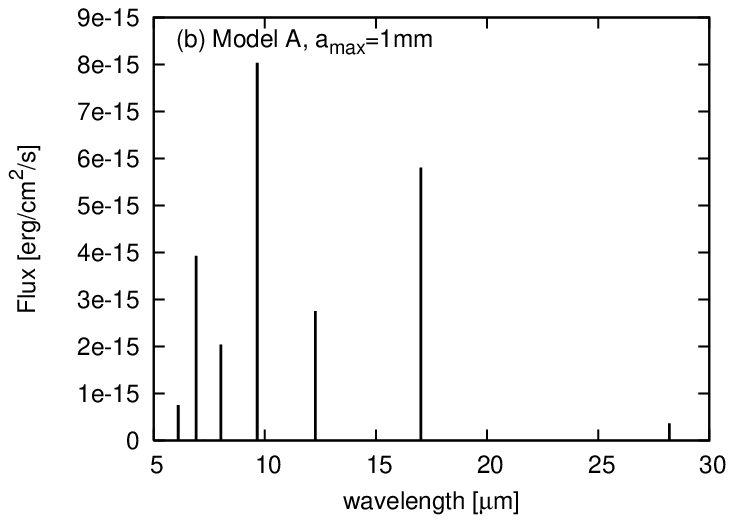}

\includegraphics[scale=1.0]{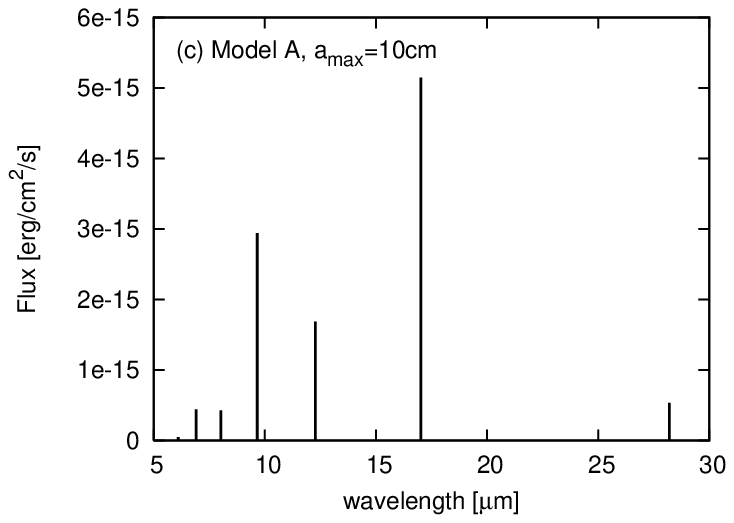}
\includegraphics[scale=1.0]{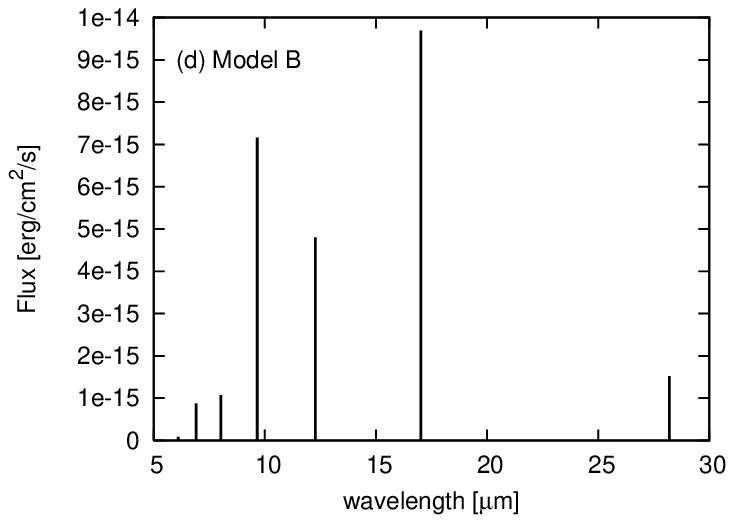}
\caption{Same as Fig.~\ref{f14}, but for the mid-infrared 
 ($5\micron <\lambda <30\micron$) spectra of pure rotational emission lines.
 The lines from lower energy levels become relatively stronger as 
 dust particles evolve and the gas temperature decreases.
 \label{f15}}
\end{figure}

\begin{figure}
\includegraphics[scale=1.0]{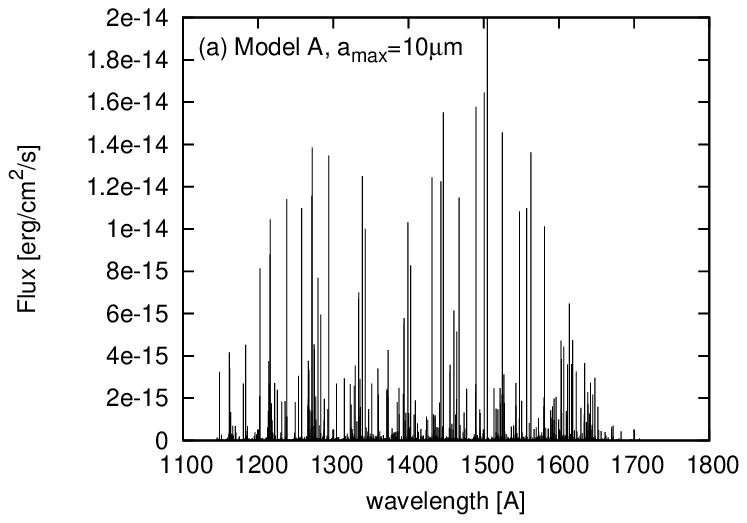}
\includegraphics[scale=1.0]{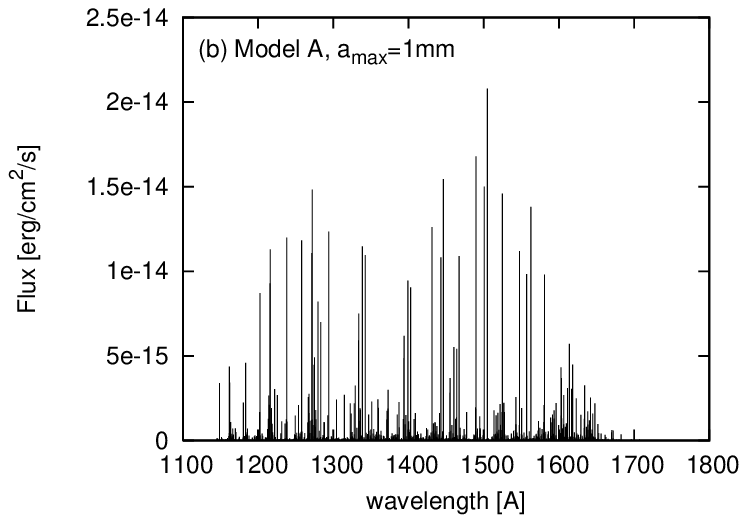}

\includegraphics[scale=1.0]{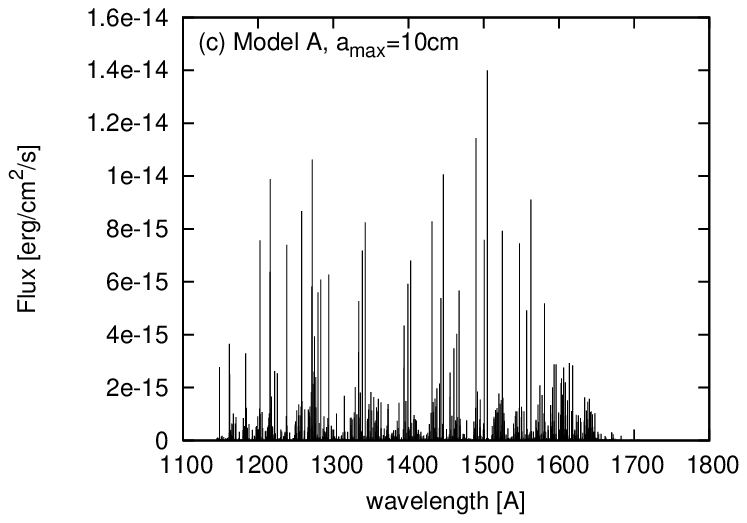}
\includegraphics[scale=1.0]{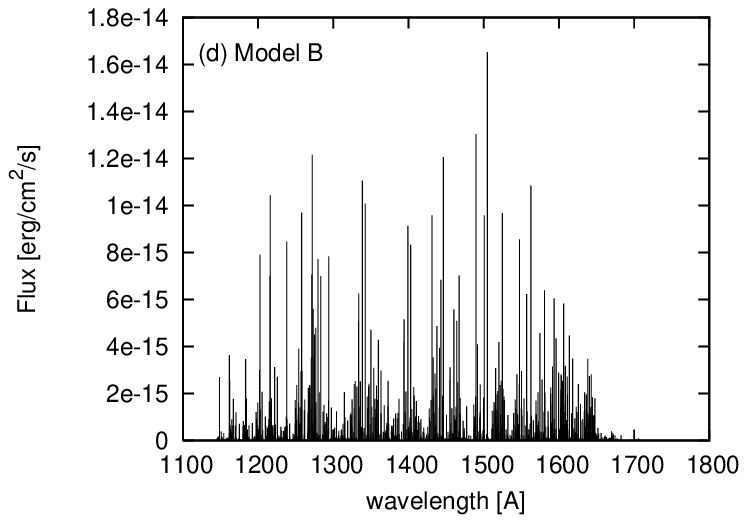}
\caption{Same as Fig.~\ref{f14}, but for the ultraviolet (1100\AA
 $<\lambda <$ 1800\AA) emission lines.
 The line fluxes of transitions originally pumped from
higher energy levels seems to be relatively weaker as dust particles
evolve and the gas temperature decreases. The lines are also affected by
 the strength of UV radiation field in the disk. \label{f16}}
\end{figure}

Figures \ref{f14}, \ref{f15}, and \ref{f16} show that the
resulting line spectra in the NIR, MIR, and UV wavelength bands,
respectively. Dust model A with $a_{\rm max}=$ (a) $10\micron$, (b)
1mm, and (c) 10cm, 
and (d) model B in a quiescent disk are used in these figures. 
The spectra from a turbulent disk are not plotted in these figures as they
are almost identical to those in a quiescent disk.

In Figure \ref{f14} and Table \ref{T1} (upper rows) we present the
ro-vibrational line fluxes of molecular hydrogen in the NIR
wavelength band. These show that as the dust particles grow, the
transition lines from higher vibrational energy levels become
relatively 
stronger in this wavelength band. This is because the level populations
change from LTE to non-LTE distributions due to the decrease of the gas
temperature and the increasing importance of UV pumping, 
so that the populations in
higher vibrational energy levels become relatively larger as we have
seen in the previous section (see Fig.~\ref{f13}). The line fluxes
decrease with the dust evolution because the area of high temperature
region in the disk shrinks. 
In Figure \ref{f15} and Table \ref{T1} (lower rows) we present the pure
rotational transition lines in the MIR wavelength band. In
this case the transition lines from lower energy levels become
relatively stronger. This is because the level populations in the ground
vibrational state are in LTE for all dust models, and
the populations in lower energy levels become relatively larger as the 
gas temperature decreases with increasing dust size. The line fluxes
decrease with dust evolution for the same reason as the NIR lines. The MIR
flux for dust model B, however, does not decrease so much because
$f_{\rm dust}$ is not very small in the outer disk (see Fig.~\ref{f2}a).
The line fluxes from lower energy levels are rather stronger than
those in model A since the gas temperatures near the midplane, to
which the MIR lines are sensitive, are higher for model B (see
\S\ref{S3.2.2}). 
In Table \ref{T1} we also list the infrared line fluxes for different
irradiation models, calculated by using dust model A with $a_{\rm
max}=10\micron$. For the irradiation model of X-rays only, the intensity
of NIR lines is similar to that for the irradiation model of X-rays $+$
UV and dust model A with $a_{\rm max}=1$mm. The MIR lines are 
relatively weaker because the gas temperature at the outer disk is not
so high if the irradiation source is X-rays only (see Fig.~\ref{f5}).
For the irradiation model of UV only, the NIR and MIR line fluxes are a bit
weaker than those for the irradiation model of X-rays $+$ UV and the dust
model A with $a_{\rm max}=10\micron$ since the gas temperature for the
latter model is higher due to the X-ray heating. The intensity of
emission lines from lower energy levels in the MIR is similar between
the former and the latter models because X-ray heating does
not affect the gas temperature in the outer disk.

Comparing our results for the 2.12 $\micron\ v=1\rightarrow 0\ S(1)$
transition to the 
observational data towards TW Hya of $1.0\times 10^{-15}$ergs s$^{-1}$
cm$^{-2}$ (Bary et al. 2003), dust model A with $a_{\rm max}=1$mm
($f_{\rm dust}=0.1$) 
seems to be most suitable. We may need larger amount of small dust
grains than that predicted in dust model B in order to reproduce the
observed 2.12 $\micron$ line flux. The calculated fluxes of the MIR
lines are consistent with the upper limits of the ground based
observations 

\onecolumn

\twocolumn

\noindent
(Richter et al. 2002; Sako et al. 2005). 

In Figure \ref{f16} and Table \ref{T2} we present the line fluxes
in the UV wavelength band. In the table we list the
lines pumped by $0-2\ R(0)$, $0-2\ R(1)$, $1-2\ P(5)$, $1-2\ R(6)$,
$3-1\ P(14)$, and $4-3\ P(5)$ transitions in the wavelength band of
the strong Ly $\alpha$ emission from the central star (see e.g., Herczeg
et al. 2002; Paper I). The figure and table show that the fluxes of
most lines decrease as dust particles evolve in the disk and the
area of the high temperature region shrinks. The fluxes of some lines,
however, do not decrease because the lines in the UV wavelength band,
excited by UV photons, are 
affected very much by the strength of the UV radiation field in the 
disk, which becomes stronger with dust evolution (see Fig.~\ref{f11}). 
The line fluxes for dust model B are relatively strong since the
populations of 
molecular hydrogen in the energy levels in the ground electronic states,
from which the lines are pumped, are larger. The populations in lower
energy levels are large because the gas temperatures near the midplane
are high, while those in higher energy levels, which are
controlled by the UV pumping process, are large because the UV
irradiation from the central star penetrates deeper in the disk
(see Figs.~\ref{f8} and \ref{f11}). We note that in this work the fluxes
of lines in the UV wavelength band caused by X-ray pumping are not
calculated. Such pumping will affect the strength of weak emission lines 
of molecular hydrogen (e.g., Bergin et al. 2004), but the effect on
the flux of the strong lines, listed in Table \ref{T2}, which are pumped
by the strong Ly $\alpha$ line emission, will be negligible. 

%
The calculated UV line fluxes are $10^{-15}\sim 10^{-14}$ ergs
s$^{-1}$ cm$^{-2}$, which are consistent with the observations towards
TW Hya (Herczeg et al. 2002) to
zeroth order but have discrepancies in details. This could be because the
UV line fluxes depend not only on the density and temperature profiles of
protoplanetary disks, but also on the strength and shape of the Ly
$\alpha$ 
line irradiated from the central star, as shown by Herczeg et al. (2002,
2004; see also Paper I). A simple single Gaussian profile is used for
the Ly $\alpha$ line profile in this paper, while the actual line
profile seems to be influenced by wind absorption (see Herczeg et
al. 2002, 2004). Thus, a more detailed analysis of the Ly $\alpha$ line
profile, beyond the scope of this work, will be needed
in order to fit the observed UV line fluxes in more detail.

Our results show that molecular hydrogen emission is strong and will be
easier to observe toward those disks whose central stars have strong UV and
X-ray radiation. In addition, if the disk contains relatively large
amount of small dust grains, the volume of hot gas in the disk will be
larger, and emission lines will be stronger. Therefore, the disks which
have an observable signature of the presence of small dust grains, 
such as strong 10$\micron$ silicate emission, could be good targets in
which to observe molecular hydrogen lines.

\subsubsection{$v=1\rightarrow 0\ S(1)/v=2\rightarrow 1\ S(1)$ Line Ratio}

\begin{figure}[t]
\includegraphics[scale=0.85]{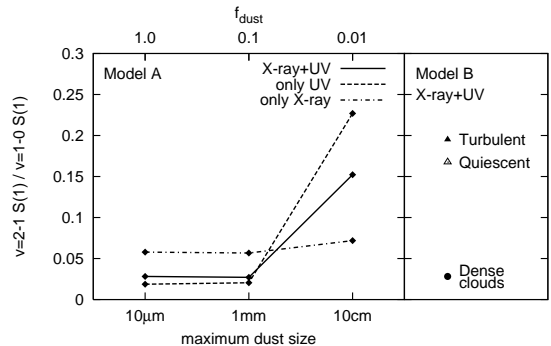}
\caption{The $v=1\rightarrow 0\ S(1)/v=2\rightarrow 1\ S(1)$ line ratios
 for the irradiation models of X-rays $+$ UV ({\it solid lines}), 
 UV only ({\it dashed lines}), and X-rays only ({\it dot-dashed lines}) and
 dust model A with $a_{\rm max}=10\micron$, 1mm, and 10cm 
 ($f_{\rm dust}=1.0$, 0.1, and 0.01) (in the left-hand panel). The
 ratios for the 
 irradiation model by X-rays $+$ UV with dust model B in quiescent
 ({\it open triangle}) and turbulent ({\it filled triangle}) disks and
 the dense clouds (dust model A with 
 $a_{\rm max}=10\micron$; {\it filled circle}) are plotted in the
 right-hand panel. If the disk is irradiated by strong UV radiation, the 
 line ratio
 increases with dust evolution as a consequence of
 the decrease in gas temperature and the increase in the UV
 radiation field strength .
\label{f17}}
\end{figure}

Finally, we discuss the effect of the dust evolution on a particular
line ratio, the $v=1\rightarrow 0\ S(1)/v=2\rightarrow 1\ S(1)$ ratio,
which is 
often used as a probe of physical properties of astronomical objects.
In Figure \ref{f17} the resulting line ratios for various irradiation
models and dust models are plotted. The lines with diamonds in the left
hand side of the figure show the ratios calculated using dust
model A with $a_{\rm max}=10\mu$m, 1mm, and 10cm ($f_{\rm dust}=1.0$,
0.1, and 0.01) and irradiation of both X-rays and UV
({\it solid lines}), UV only ({\it dashed
lines}), and X-rays only ({\it dot-dashed lines}).
They show that if the disk is irradiated by the UV radiation from
the central star, this ratio becomes larger as the dust particles grow
and the total surface area of dust grains, $f_{\rm dust}$, decreases.
This is because the level populations
of molecular hydrogen change from LTE to non-LTE distributions due to
the increase in the grain photoelectric heating rate and the decrease of
grain opacity, as discussed in \S\ref{S3.3.2}. This effect appears to be
more efficient if the disk is irradiated only by UV and the
heating source of the gas at the surface disk is grain
photoelectric heating only. If the disk is irradiated by X-rays
only, the ratio does not change with the maximum dust size. 
We note, however, that we have neglected the change in the X-ray
photoionization cross section due to the dust evolution. 
If we take it into account, it affects the line ratio
slightly with the error being about 25\% for the model with $a_{\rm
max}=10$cm in the extreme case that the contribution to the cross
sections of heavy elements in the dust is negligible (see
\S\ref{S3.2.2}; Wilms et al. 2000). 
Figure \ref{f17} shows that the line ratio in the model with
X-ray irradiation only is slightly larger than that in the model with
UV irradiation, 
although the gas temperature is lower. This occurs because the level
populations are not in LTE, but affected by the X-ray pumping
process (see \S\ref{S3.2.1}). 
We also plot in the right hand side of
Figure \ref{f17} the resulting line ratios for dust model B in
quiescent and turbulent disks in filled and open triangles,
respectively. The ratio for the 
model with dust grains typical of dense cloud (model A with $a_{\rm
max}=10\mu$m) is plotted as a filled circle. The results suggest that if
the dust grains coagulate and settle towards the disk midplane, the
ratio becomes substantially larger than the case in which dust grains do
not evolve. So, our results suggest that dust evolution
in protoplanetary disks could be observable through this particular line
ratio. Itoh et al. (2003) derived an upper
limit to the line ratio of 0.26 from an observation toward a classical T
Tauri star, LkH$\alpha$ 264, and Bary et al. (2007, in preparation) find
an upper limit of $\sim$0.2 toward a Herbig Be star, HD 97048; all the
models we used in this paper almost satisfy these observational upper
limits.  

\section{Summary}

We have made a detailed model of physical structure of protoplanetary
disks and calculated the level populations and line emission of molecular
hydrogen, taking into account X-ray irradiation from the central star
as well as dust growth and settling towards the disk midplane.
%
%
We have followed the time evolution of the spatial and size
distributions of dust particles in the disks by numerically solving the
coagulation equation for settling dust particles. The resulting 
mass and total surface area of dust grains per unit gas volume is much
smaller, except at the disk midplane, than those for the model in which 
the dense cloud dust grains 
are well mixed with the gas. At the disk surface the dust density
normalized by the initial value, $\rho_{\rm dust}/\rho_{{\rm dust}, 0}$,  
and the parameter for the total surface area of dust grains, $f_{\rm
dust}$, are small due to the dust settling toward the
disk midplane. Near the disk midplane the parameter $f_{\rm 
dust}$ becomes much smaller since small dust particles are removed by 
dust coagulation.

We have studied the effects of X-ray irradiation on the physical structure 
of the disks and found 
that the X-ray irradiation is the dominant heating source in the inner
region and in the surface layers of the disk. FUV
heating dominates in the middle layers and in the outer region of the disk.
This is because the FUV radiation is scattered efficiently by dust
grains, while the Compton scattering of X-ray radiation is inefficient
in the energy range of $E\la 1$keV in which T Tauri 
stars mainly emit X-ray radiation. We found that the ionization
rate caused by X-rays is much smaller than that due to interstellar 
cosmic-rays near the disk midplane because of the relatively large 
attenuation and inefficient scattering of X-rays.

The dust evolution in the disks affects the physical disk structure,
especially the gas temperature at the disk surface and the FUV
radiation field within the disk. As the dust particles grow or settle
towards the disk midplane, the gas temperature in the middle layers and
the outer disk decreases because the grain photoelectric
heating which is induced by FUV radiation becomes less efficient. 
Meanwhile, the
FUV radiation from the central star penetrates deeper into the disk due
to the decrease of grain opacity.

Furthermore, making use of the obtained physical structure of the disks,
we calculated the 
level populations of molecular hydrogen in the ground electronic state. 
Our results show that if the central star has strong X-ray and weak UV
radiation, the level populations are controlled by X-ray pumping. 
Otherwise, the level populations are mainly
controlled by thermal collisions or UV pumping, depending on the
dust properties in the disk. As the dust particles evolve in the disk,
the level populations change from LTE to non-LTE distributions since 
collisional excitation becomes less efficient due to
the decrease of the gas temperature at the disk surface while UV
pumping becomes more efficient owing to the stronger UV
radiation field in the disk.

Finally, using these level populations, we calculated the line emission of
molecular hydrogen from the disk. The ro-vibrational line spectra
in the near-infrared wavelength band show that the emission lines from high
energy levels become relatively stronger as the dust particles evolve
in the disk. Again, this is due to level populations changing from LTE to
non-LTE and the populations in higher vibrational energy
levels becoming relatively larger. For the pure rotational
line spectra in the mid-infrared wavelength band, it is the emission lines
from lower energy levels which become relatively stronger. This is because the
level populations in the ground vibrational state
are in LTE and the populations in lower energy levels
become relatively larger with dust evolution and
decreasing dust surface area which results in lower gas temperatures. 
For transitions in the UV wavelength band,
the dependence on the dust evolution is not so straightforward because
the line fluxes decrease 
as the area of high temperature region shrinks, while they increase as
UV irradiation from the central star penetrates deeper in the
disks. Basically, the line fluxes which originate from pumping from
higher energy levels seem to be relatively weaker as the dust particles
evolve and the gas temperature decreases.
Our results suggest that 
infrared line ratios of molecular hydrogen could be a useful probe 
of dust evolution in protoplanetary disks. 
If the dust particles evolve, the $v=1\rightarrow
0\ S(1)/v=2\rightarrow 1\ S(1)$ line ratio, for example, becomes clearly
larger than that for the dense cloud dust model
(without the dust evolution). Further observations of the line
ratios of molecular hydrogen could provide some constraints on the dust
evolution model in protoplanetary disks.

\acknowledgments

We are grateful to an anonymous referee for his comments which
improve the clarity of our discussion.
We would like to thank J.S. Bary, D.A. Weintraub, M.J. Richter, and
M.A. Bitner for giving us information about unpublished observational
data, Y. Itoh and T. Takeuchi for fruitful comments,
and T. Matsuda for useful help for numerical calculations.
This work is supported by ``The 21st Century COE Program of Origin and
Evolution of Planetary Systems" and the Grants-in-Aid for
Scientific Research 17039008, 17540217, and 18026006 in MEXT. 
Astrophysics at Queen's University Belfast is supported by PPARC.
H.\ N. and M.\,T. acknowledge financial supported from the Japan Society
for the Promotion of Science.

\clearpage


\begin{deluxetable}{cc|cc|ccc|cc}
\tabletypesize{\scriptsize}
\tablecaption{Calculated infrared line fluxes of molecular hydrogen.\label{T1}}
\tablewidth{0pt}
\tablehead{
   & & \multicolumn{7}{c}{$F_{\rm line}$ ($\times 10^{-15}$ erg/s/cm$^2$)} \\
 $\lambda$ & Transition & X-rays & UV & \multicolumn{5}{c}{X-rays + UV} \\
 ($\mu$m) & & \multicolumn{5}{|c|}{Model A ($a_{\rm max}$)} & \multicolumn{2}{|c}{Model B} \\
 & & 10$\micron$ & 10$\micron$ & 10$\micron$ & 1mm & 10cm & Quiescent & Turbulent \\
}
\startdata
  1.16 & 2-0\ S(1) &  0.068 &  0.182 &  0.439 &  0.051 &  0.016 &  0.020 &  0.019 \\
  1.19 & 2-0\ S(0) &  0.013 &  0.039 &  0.091 &  0.013 &  0.005 &  0.006 &  0.006 \\
  1.24 & 2-0\ Q(1) &  0.043 &  0.143 &  0.310 &  0.056 &  0.057 &  0.067 &  0.061 \\
  1.24 & 2-0\ Q(2) &  0.014 &  0.041 &  0.095 &  0.014 &  0.008 &  0.009 &  0.009 \\
  1.25 & 2-0\ Q(3) &  0.043 &  0.116 &  0.280 &  0.033 &  0.010 &  0.014 &  0.013 \\
  1.83 & 1-0\ S(5) &  0.587 &  2.286 &  4.849 &  0.446 &  0.029 &  0.050 &  0.049 \\
  1.89 & 1-0\ S(4) &  0.314 &  1.555 &  3.007 &  0.287 &  0.017 &  0.027 &  0.026 \\
  1.96 & 1-0\ S(3) &  1.326 &  8.096 & 14.475 &  1.472 &  0.079 &  0.119 &  0.111 \\
  2.03 & 1-0\ S(2) &  0.533 &  3.857 &  6.477 &  0.719 &  0.041 &  0.058 &  0.051 \\
  2.12 & 1-0\ S(1)\tablenotemark{a} &  1.565 & 12.968 & 20.725 &  2.526 &  0.142 &  0.169 &  0.135 \\
  2.22 & 1-0\ S(0) &  0.357 &  3.256 &  5.023 &  0.667 &  0.040 &  0.051 &  0.037 \\
  2.25 & 2-1\ S(1) &  0.090 &  0.242 &  0.583 &  0.068 &  0.022 &  0.028 &  0.026 \\
  2.35 & 2-1\ S(0) &  0.019 &  0.056 &  0.130 &  0.019 &  0.011 &  0.013 &  0.012 \\
  2.41 & 1-0\ Q(1) &  1.270 & 12.299 & 18.528 &  2.707 &  0.355 &  0.484 &  0.373 \\
  2.41 & 1-0\ Q(2) &  0.395 &  3.604 &  5.560 &  0.748 &  0.066 &  0.100 &  0.077 \\
  2.42 & 1-0\ Q(3) &  1.101 &  9.128 & 14.588 &  1.780 &  0.117 &  0.185 &  0.145 \\
  2.44 & 1-0\ Q(4) &  0.298 &  2.156 &  3.621 &  0.402 &  0.023 &  0.034 &  0.029 \\
  2.45 & 1-0\ Q(5) &  0.648 &  3.954 &  7.069 &  0.719 &  0.039 &  0.059 &  0.055 \\
  2.47 & 1-0\ Q(6) &  0.142 &  0.702 &  1.357 &  0.129 &  0.008 &  0.012 &  0.012 \\
  2.50 & 1-0\ Q(7) &  0.260 &  1.011 &  2.145 &  0.197 &  0.013 &  0.022 &  0.021 \\
  2.55 & 2-1\ Q(1) &  0.066 &  0.223 &  0.483 &  0.088 &  0.090 &  0.110 &  0.097 \\
  2.56 & 2-1\ Q(2) &  0.022 &  0.063 &  0.147 &  0.021 &  0.012 &  0.014 &  0.013 \\
  2.57 & 2-1\ Q(3) &  0.066 &  0.176 &  0.424 &  0.050 &  0.016 &  0.021 &  0.019 \\ \tableline
  6.11 & 0-0\ S(6) &  0.368 &  3.949 &  5.791 &  0.755 &  0.050 &  0.084 &  0.058 \\
  6.91 & 0-0\ S(5) &  1.203 & 17.346 & 23.219 &  3.928 &  0.441 &  0.880 &  0.564 \\
  8.02 & 0-0\ S(4) &  0.375 &  7.023 &  8.753 &  2.038 &  0.427 &  1.072 &  0.731 \\
  9.66 & 0-0\ S(3) &  0.849 & 19.835 & 23.286 &  8.035 &  2.943 &  7.160 &  6.140 \\
 12.27 & 0-0\ S(2)\tablenotemark{b} &  0.163 &  4.271 &  4.719 &  2.751 &  1.688 &  4.803 &  4.711 \\
 17.02 & 0-0\ S(1)\tablenotemark{b,c} &  0.209 &  4.466 &  4.606 &  5.806 &  5.148 &  9.690 &  9.900 \\
 28.20 & 0-0\ S(0)\tablenotemark{c} &  0.020 &  0.145 &  0.142 &  0.365 &  0.532 &  1.521 &  1.717 \\
\enddata
\tablenotetext{a}{Ground-based observations by Bary et al. (2003) and Itoh
 et al. (2003) give fluxes of $(1.0-15)\times 10^{-15}$ erg/s/cm$^2$.}
\tablenotetext{b}{Ground-based observations by Richter et al. (2002) and
 Sako et al. (2005) give fluxes of $<30 \times 10^{-15}$ erg/s/cm$^2$
 for the 12$\micron$ S(2) line and $<39 \times 10^{-15}$ erg/s/cm$^2$
 for the 17$\micron$ S(1) line.}
\tablenotetext{c}{ISO observations by Thi et al. (2001) give fluxes of
 $(28-81) \times 10^{-15}$ erg/s/cm$^2$ for the 17$\micron$ S(1) line
 and $(25-57) \times 10^{-15}$ erg/s/cm$^2$ for the 28$\micron$ S(0)
 line.} 
\end{deluxetable}

\clearpage


\begin{deluxetable}{cc|ccc|cc|c||cc|ccc|cc|c}
\tabletypesize{\scriptsize}
\rotate
\tablecaption{Calculated and observed ultraviolet line fluxes of
 molecular hydrogen.\label{T2}}
\tablewidth{0pt}
\tablehead{
 & & \multicolumn{6}{c||}{$F_{\rm line}$ ($\times 10^{-15}$ erg/s/cm$^2$)} &
 & & \multicolumn{6}{c}{$F_{\rm line}$ ($\times 10^{-15}$ erg/s/cm$^2$)} \\
 $\lambda$ & Transition & \multicolumn{3}{|c|}{Model A ($a_{\rm max}$)} &
 \multicolumn{2}{|c|}{Model B} & Obs.\tablenotemark{a} &
 $\lambda$ & Transition & \multicolumn{3}{|c|}{Model A ($a_{\rm max}$)} &
 \multicolumn{2}{|c|}{Model B} & Obs.\tablenotemark{a} \\
 (\AA) & & 10$\micron$ & 1mm & 10cm & Quies. & Turb. & &
 (\AA) & & 10$\micron$ & 1mm & 10cm & Quies. & Turb. &
}
\startdata
 \multicolumn{8}{c||}{Pumped by 0-2\ R(0)} & \multicolumn{8}{|c}{Pumped by 0-2\ R(1)} \\
\tableline
  1161.7 & 0-1\ R(0) &  0.436 &  0.427 &  0.291 &  0.311 &  0.300 &  &  1162.2 & 0-1\ R(1) &  0.501 &  0.512 &  0.412 &  0.428 &  0.424 &  \\
  1166.3 & 0-1\ P(2) &  0.688 &  0.700 &  0.608 &  0.638 &  0.621 &  &  1169.8 & 0-1\ P(3) &  0.681 &  0.721 &  0.637 &  0.642 &  0.633 &  \\
  1217.3 & 0-2\ R(0) &  1.728 &  1.830 &  1.245 &  1.505 &  1.585 &  &  1217.7 & 0-2\ R(1) &  1.759 &  1.909 &  1.658 &  1.786 &  1.806 & 9.1 \\
  1222.0 & 0-2\ P(2) &  2.719 &  3.039 &  2.623 &  3.121 &  3.398 &  &  1225.6 & 0-2\ P(3) &  2.395 &  2.683 &  2.538 &  2.715 &  2.789 &  \\
  1274.6 & 0-3\ R(0) &  4.550 &  4.478 &  2.608 &  3.630 &  3.954 & 27.4 &  1275.0 & 0-3\ R(1) &  4.304 &  4.918 &  3.941 &  4.506 &  4.624 & 24.6 \\
  1279.6 & 0-3\ P(2) &  7.690 &  8.210 &  5.606 &  7.717 &  8.722 & 39.2 &  1283.2 & 0-3\ P(3) &  5.950 &  6.996 &  6.092 &  7.000 &  7.443 & 28.0 \\
  1333.6 & 0-4\ R(0) &  6.694 &  5.909 &  3.344 &  5.087 &  5.683 & 42.8 &  1333.9 & 0-4\ R(1) &  7.005 &  7.487 &  5.268 &  6.252 &  6.606 & 7.9 \\
  1338.7 & 0-4\ P(2) & 12.495 & 11.472 &  7.185 & 11.063 & 12.584 & 73.1 &  1342.4 & 0-4\ P(3) & 10.007 & 10.958 &  8.251 & 10.080 & 11.168 & 64.9 \\
  1393.9 & 0-5\ R(0) &  5.516 &  4.866 &  2.757 &  4.199 &  4.693 & 35.3 &  1394.1 & 0-5\ R(1) &  5.790 &  6.177 &  4.344 &  5.159 &  5.456 & 52.4 \\
  1399.1 & 0-5\ P(2) & 10.308 &  9.453 &  5.923 &  9.136 & 10.389 & 73.8 &  1402.8 & 0-5\ P(3) &  8.277 &  9.047 &  6.807 &  8.328 &  9.235 & 73.1 \\
  1455.0 & 0-6\ R(0) &  3.232 &  2.819 &  1.633 &  2.563 &  2.873 & 20.8 &  1455.2 & 0-6\ R(1) &  3.574 &  3.686 &  2.570 &  3.111 &  3.349 & 30.9 \\
  1460.4 & 0-6\ P(2) &  6.141 &  5.507 &  3.483 &  5.576 &  6.290 & 41.6 &  1464.0 & 0-6\ P(3) &  5.148 &  5.420 &  4.030 &  5.091 &  5.744 & 42.1 \\
  1516.4 & 0-7\ R(0) &  1.273 &  1.097 &  0.666 &  1.107 &  1.230 & 21.2 &  1516.5 & 0-7\ R(1) &  1.514 &  1.491 &  1.044 &  1.325 &  1.471 & 21.2 \\
  1521.8 & 0-7\ P(2) &  2.467 &  2.155 &  1.383 &  2.355 &  2.597 & 16.2 &  1525.4 & 0-7\ P(3) &  2.202 &  2.197 &  1.617 &  2.194 &  2.515 & 17.9 \\
  1577.3 & 0-8\ R(0) &  0.302 &  0.258 &  0.164 &  0.291 &  0.315 &  &  1577.1 & 0-8\ R(1) &  0.378 &  0.361 &  0.259 &  0.362 &  0.412 &  \\
  1582.6 & 0-8\ P(2) &  0.592 &  0.509 &  0.330 &  0.593 &  0.637 &  &  1586.0 & 0-8\ P(3) &  0.553 &  0.531 &  0.389 &  0.590 &  0.662 &  \\
  1636.4 & 0-9\ R(0) &  0.034 &  0.029 &  0.019 &  0.035 &  0.037 &  &  1636.1 & 0-9\ R(1) &  0.044 &  0.041 &  0.030 &  0.048 &  0.053 &  \\
  1641.7 & 0-9\ P(2) &  0.067 &  0.058 &  0.038 &  0.069 &  0.073 &  &  1644.8 & 0-9\ P(3) &  0.064 &  0.061 &  0.044 &  0.073 &  0.079 &  \\
  1692.5 & 0-10\ R(0) &  0.001 &  0.001 &  0.000 &  0.001 &  0.001 &  &  1692.0 & 0-10\ R(1) &  0.001 &  0.001 &  0.001 &  0.001 &  0.001 &  \\
  1697.6 & 0-10\ P(2) &  0.002 &  0.001 &  0.001 &  0.002 &  0.002 &  & 1700.3 & 0-10\ P(3) &  0.002 &  0.001 &  0.001 &  0.002 &  0.002 &  \\ \tableline
 \multicolumn{8}{c||}{Pumped by 1-2\ P(5)} & \multicolumn{8}{|c}{Pumped by 1-2\ R(6)} \\
\tableline
  1148.7 & 1-1\ R(3) &  3.247 &  3.386 &  2.774 &  2.701 &  2.688 & 4.6 &  1162.0 & 1-1\ R(6) &  3.417 &  3.415 &  2.505 &  2.559 &  2.525 &  \\
  1161.9 & 1-1\ P(5) &  4.165 &  4.369 &  3.660 &  3.631 &  3.611 & 10.9 &  1183.4 & 1-1\ P(8) &  4.524 &  4.591 &  3.294 &  3.467 &  3.404 &  \\
  1202.5 & 1-2\ R(3) &  8.145 &  8.716 &  7.575 &  7.909 &  7.881 & 11.3 &  1215.8 & 1-2\ R(6) &  8.795 &  9.284 &  6.379 &  6.986 &  6.787 &  \\
  1216.2 & 1-2\ P(5) & 10.449 & 11.295 &  9.889 & 10.442 & 10.425 &  &  1238.0 & 1-2\ P(8) & 11.411 & 11.984 &  7.394 &  8.460 &  8.162 & 11.5 \\
  1257.9 & 1-3\ R(3) & 10.991 & 11.831 &  8.676 &  9.699 &  9.640 & 18.1 &  1271.2 & 1-3\ R(6) & 11.569 & 11.074 &  5.822 &  7.064 &  6.754 & 14.1 \\
  1272.0 & 1-3\ P(5) & 13.849 & 14.827 & 10.630 & 12.158 & 12.127 & 20.5 &  1294.0 & 1-3\ P(8) & 13.476 & 12.341 &  6.279 &  7.831 &  7.423 & 13.0 \\
  1314.8 & 1-4\ R(3) &  2.933 &  2.700 &  1.686 &  2.061 &  2.048 & 12.2 &  1327.8 & 1-4\ R(6) &  2.581 &  2.181 &  1.062 &  1.362 &  1.270 & 6.1 \\
  1329.3 & 1-4\ P(5) &  3.536 &  3.240 &  2.022 &  2.530 &  2.489 & 7.5 &  1351.2 & 1-4\ P(8) &  2.689 &  2.303 &  1.129 &  1.437 &  1.347 & 2.8 \\
  1372.7 & 1-5\ R(3) &  2.063 &  1.892 &  1.180 &  1.449 &  1.438 & 3.2 &  1385.2 & 1-5\ R(6) &  1.813 &  1.530 &  0.745 &  0.956 &  0.891 &  \\
  1387.5 & 1-5\ P(5) &  2.486 &  2.272 &  1.416 &  1.776 &  1.745 & 7.1 &  1409.2 & 1-5\ P(8) &  1.899 &  1.621 &  0.793 &  1.011 &  0.947 & 2.2 \\
  1431.2 & 1-6\ R(3) & 12.444 & 12.615 &  8.287 &  9.576 &  9.548 & 29.0 &  1443.1 & 1-6\ R(6) & 12.251 & 10.827 &  5.388 &  6.837 &  6.427 & 11.3 \\
  1446.3 & 1-6\ P(5) & 15.518 & 15.441 & 10.068 & 12.060 & 12.049 & 44.2 &  1467.4 & 1-6\ P(8) & 11.488 & 10.895 &  5.671 &  7.012 &  6.672 & 17.6 \\
  1489.8 & 1-7\ R(3) & 15.769 & 16.791 & 11.439 & 13.034 & 12.946 & 48.2 &  1500.7 & 1-7\ R(6) & 16.461 & 15.004 &  7.589 &  9.577 &  9.034 & 19.7 \\
  1505.0 & 1-7\ P(5) & 19.978 & 20.803 & 14.002 & 16.528 & 16.532 & 57.5 &  1524.9 & 1-7\ P(8) & 14.565 & 14.593 &  7.929 &  9.677 &  9.255 & 23.5 \\

  1547.6 & 1-8\ R(3) & 10.831 & 11.200 &  7.458 &  8.565 &  8.528 & 35.3 &  1557.2 & 1-8\ R(6) & 10.993 &  9.828 &  4.920 &  6.230 &  5.864 & 17.0 \\
  1562.7 & 1-8\ P(5) & 13.631 & 13.810 &  9.112 & 10.847 & 10.845 & 37.2 &  1581.0 & 1-8\ P(8) & 10.123 &  9.813 &  5.191 &  6.384 &  6.088 & 17.5 \\
  1603.5 & 1-9\ R(3) &  3.851 &  3.685 &  2.342 &  2.774 &  2.773 & 11.2 &  1611.3 & 1-9\ R(6) &  3.608 &  3.098 &  1.520 &  1.941 &  1.816 &  \\
  1618.2 & 1-9\ P(5) &  4.742 &  4.487 &  2.843 &  3.482 &  3.459 & 11.6 &  1634.2 & 1-9\ P(8) &  3.653 &  3.256 &  1.630 &  2.048 &  1.935 & 5.5 \\
  1656.2 & 1-10\ R(3) &  0.443 &  0.402 &  0.250 &  0.312 &  0.307 &  &  1661.5 & 1-10\ R(6) &  0.395 &  0.332 &  0.161 &  0.207 &  0.193 &  \\
  1670.0 & 1-10\ P(5) &  0.537 &  0.487 &  0.303 &  0.382 &  0.374 &  &  1682.8 & 1-10\ P(8) &  0.425 &  0.359 &  0.175 &  0.224 &  0.209 &  \\ \tableline
 \multicolumn{8}{c||}{Pumped by 3-1\ P(14)} & \multicolumn{8}{|c}{Pumped by 4-3\ P(5)} \\
\tableline
  1180.3 & 3-1\ R(12) &  2.681 &  2.242 &  0.848 &  0.827 &  0.777 &  &  1151.3 & 4-2\ R(3) &  0.262 &  0.196 &  0.096 &  0.153 &  0.170 &  \\
  1214.2 & 3-1\ P(14) &  3.749 &  2.662 &  0.854 &  0.919 &  0.846 &  &  1163.8 & 4-2\ P(5) &  0.333 &  0.244 &  0.117 &  0.202 &  0.217 &  \\
  1231.3 & 3-2\ R(12) &  1.839 &  1.133 &  0.338 &  0.405 &  0.354 &  &  1202.0 & 4-3\ R(3) &  2.094 &  1.680 &  0.823 &  1.218 &  1.354 &  \\
  1266.1 & 3-2\ P(14) &  1.651 &  1.064 &  0.324 &  0.367 &  0.331 &  &  1214.9 & 4-3\ P(5) &  2.699 &  2.108 &  1.025 &  1.666 &  1.815 &  \\
  1283.1 & 3-3\ R(12) &  0.345 &  0.203 &  0.059 &  0.075 &  0.063 &  &  1253.8 & 4-4\ R(3) &  3.058 &  2.094 &  0.945 &  1.597 &  1.745 & 6.0 \\
  1318.7 & 3-3\ P(14) &  0.328 &  0.195 &  0.057 &  0.071 &  0.061 &  &  1267.0 & 4-4\ P(5) &  3.771 &  2.551 &  1.161 &  2.238 &  2.324 & 12.9 \\
  1335.4 & 3-4\ R(12) &  2.896 &  1.914 &  0.590 &  0.671 &  0.605 &  &  1306.4 & 4-5\ R(3) &  0.142 &  0.091 &  0.041 &  0.086 &  0.085 &  \\
  1371.5 & 3-4\ P(14) &  2.432 &  1.746 &  0.564 &  0.605 &  0.557 & 4.0 &  1319.8 & 4-5\ P(5) &  0.171 &  0.109 &  0.049 &  0.105 &  0.102 &  \\
  1387.7 & 3-5\ R(12) &  1.246 &  0.772 &  0.231 &  0.275 &  0.241 & 3.4 &  1359.3 & 4-6\ R(3) &  3.399 &  2.430 &  1.105 &  1.751 &  1.938 & 6.5 \\
  1423.9 & 3-5\ P(14) &  1.121 &  0.729 &  0.223 &  0.251 &  0.227 & 1.6 &  1372.9 & 4-6\ P(5) &  4.266 &  2.993 &  1.380 &  2.534 &  2.689 & 7.2 \\
  1439.4 & 3-6\ R(12) &  0.597 &  0.360 &  0.106 &  0.130 &  0.112 &  &  1412.0 & 4-7\ R(3) &  0.802 &  0.526 &  0.236 &  0.443 &  0.466 &  \\
  1475.2 & 3-6\ P(14) &  0.557 &  0.346 &  0.103 &  0.121 &  0.107 &  &  1425.6 & 4-7\ P(5) &  0.976 &  0.637 &  0.287 &  0.587 &  0.590 &  \\
  1489.4 & 3-7\ R(12) &  2.669 &  1.871 &  0.595 &  0.657 &  0.600 & 6.7 &  1463.8 & 4-8\ R(3) &  1.957 &  1.360 &  0.615 &  1.014 &  1.115 & 2.3 \\
  1524.3 & 3-7\ P(14) &  2.166 &  1.680 &  0.573 &  0.599 &  0.555 & 5.1 &  1477.3 & 4-8\ P(5) &  2.440 &  1.671 &  0.764 &  1.446 &  1.515 & 5.8 \\
  1536.8 & 3-8\ R(12) &  0.658 &  0.403 &  0.120 &  0.144 &  0.126 &  &  1513.7 & 4-9\ R(3) &  2.463 &  1.787 &  0.815 &  1.269 &  1.408 & 4.3 \\
  1569.8 & 3-8\ P(14) &  0.612 &  0.390 &  0.118 &  0.135 &  0.121 &  &  1526.8 & 4-9\ P(5) &  3.125 &  2.221 &  1.028 &  1.861 &  1.986 & 6.1 \\
  1580.0 & 3-9\ R(12) &  1.616 &  1.082 &  0.335 &  0.379 &  0.343 & 8.9 &  1560.6 & 4-10\ R(3) &  0.819 &  0.546 &  0.245 &  0.440 &  0.472 &  \\
  1609.9 & 3-9\ P(14) &  1.385 &  1.012 &  0.331 &  0.352 &  0.325 & 8.5 &  1572.9 & 4-10\ P(5) &  1.011 &  0.668 &  0.302 &  0.604 &  0.615 & 6.8 \\
  1617.0 & 3-10\ R(12) &  3.614 &  3.050 &  1.108 &  1.154 &  1.075 & 6.8 &  1602.8 & 4-11\ R(3) &  4.721 &  4.325 &  2.165 &  2.803 &  3.057 & 17.6 \\
  1642.1 & 3-10\ P(14) &  2.722 &  2.538 &  1.069 &  1.068 &  1.001 &  &  1613.9 & 4-11\ P(5) &  6.466 &  5.715 &  2.925 &  4.464 &  4.952 & 20.4 \\
\enddata
\tablenotetext{a}{FUSE and HST observations by Herczeg et al. (2002).}
\end{deluxetable}

\end{document}